\documentclass[journal]{IEEEtran}
\usepackage{graphicx,amssymb,amsmath}
\usepackage[noadjust]{cite}
\usepackage{setspace}               
\usepackage{stfloats}
\usepackage{bbding}
\usepackage{amsthm}
\usepackage{amsmath}
\usepackage{cases,subeqnarray}
\usepackage{bm,multirow,bigstrut}
\usepackage{epstopdf}
\usepackage{textcomp}
\usepackage{latexsym,bm}
\usepackage{booktabs,changebar}
\usepackage{xcolor}
\usepackage{mathtools}
\usepackage{dsfont}
\usepackage{extarrows}
\usepackage{mathrsfs}
\usepackage{subfigure}

\usepackage{cite}
\usepackage{bm}
\usepackage{cleveref}
\usepackage{multicol}       
\usepackage{multirow}       
\usepackage{array}          
\usepackage{colortbl}
\usepackage{makecell}
\definecolor{crimson}{RGB}{192,0,0}         
\definecolor{navy}{RGB}{47,85,151}         

\makeatletter                               
\newif\if@restonecol
\makeatother

\makeatletter
\newif\if@restonecol
\makeatother

\usepackage[linesnumbered,ruled,vlined]{algorithm2e}
\usepackage{algpseudocode}
\usepackage{amsmath}
\renewcommand{\arraystretch}{1.5} %

\theoremstyle{plain}
\newtheorem{thm}{Theorem}
\newtheorem{lemm}{Lemma}

\newtheorem{coro}{Corollary}

\theoremstyle{plain}
\newtheorem{rem}{Remark}

\IEEEoverridecommandlockouts

\begin{document}
\title{Robust Multidimensional Graph Neural Networks for Signal Processing in Wireless Communications with Edge-Graph Information Bottleneck}
\author{{Ziheng~Liu,~\IEEEmembership{Student Member,~IEEE}, Jiayi~Zhang,~\IEEEmembership{Senior Member,~IEEE}, Yiyang~Zhu,~\IEEEmembership{Student Member,~IEEE}, Enyu~Shi,~\IEEEmembership{Student Member,~IEEE}, and Bo~Ai,~\IEEEmembership{Fellow,~IEEE}}
\thanks{Z. Liu, J. Zhang, Y. Zhu, E. Shi, and B. Ai are with the State Key Laboratory of Advanced Rail Autonomous Operation, and also with School of Electronic and Information Engineering, Beijing Jiaotong University, Beijing 100044, China (e-mail: 23111013@bjtu.edu.cn; jiayizhang@bjtu.edu.cn; 21251058@bjtu.edu.cn; 21111047@bjtu.edu.cn; boai@bjtu.edu.cn).}}
\maketitle
\begin{abstract}
Signal processing is crucial for satisfying the high data rate requirements of future sixth-generation (6G) wireless networks. However, the rapid growth of wireless networks has brought about massive data traffic, which hinders the application of traditional optimization theory-based algorithms.
Meanwhile, traditional graph neural networks (GNNs) focus on compressing inputs onto vertices to update representations, which often leads to their inability to effectively distinguish input features and severely weakens performance. In this context, designing efficient signal processing frameworks becomes imperative. Moreover, actual scenarios are susceptible to multipath interference and noise, resulting in specific differences between the received and actual information. To address these challenges, this paper incorporates multidimensional graph neural networks (MDGNNs) with edge-graph information bottleneck (EGIB) to design a robust framework for signal processing. Specifically, MDGNNs utilize hyper-edges instead of vertices to update representations to avoid indistinguishable features and reduce information loss, while EGIB encourages providing minimal sufficient information about outputs to avoid aggregation of irrelevant information. We numerically demonstrate that compared with existing frameworks, the proposed frameworks achieve excellent performance in terms of spectrum efficiency (SE) and network overhead under multiple signal processing tasks. Remarkably, as the interference noise increases, the SE performance of the proposed frameworks gradually stabilizes. This reveals the proposed frameworks have excellent robustness in interference prone environments, especially in wireless policies related to channel matrices.
\end{abstract}
\begin{IEEEkeywords}
Edge-graph information bottleneck, multidimensional graph neural networks, robust, signal processing, wireless communications.
\end{IEEEkeywords}
\IEEEpeerreviewmaketitle
\newcounter{mytempeqncnt_1}
\vspace{-0.3cm}
\section{Introduction}
\subsection{Motivation of Multidimensional Graph Neural Networks}
Signal processing has received significant attention in wireless communication since they can dramatically improve system performance through signal enhancement and interference management, such as joint precoding and power control techniques \cite{[36],[1],[2]}. Meanwhile, the continuous innovation and optimization of signal processing techniques are driving the sustained development of wireless communications \cite{[3]}, laying the foundation for the advancement of future sixth-generation (6G) wireless networks to meet substantially higher performance requirements \cite{[4],[5],[6],[44]}. However, owing to the massive amount of data traffic brought about by the Internet of Everything (IoE) \cite{[7]}, traditional algorithms that rely on optimization theory face the challenges of sizeable dimensional matrix inversion and extensive iteration, such as weighted minimum mean-square error (WMMSE) \cite{[8],[9]}, which hinders their application in real-time communication systems.

Recently, machine learning (ML) has been envisioned as an essential technology for promoting the development of 6G networks \cite{[3],[10]}, say in achieving good performance with lower computational complexity and inference time, which can better handle computationally intensive and time-sensitive signal processing tasks \cite{[11],[12]}. The main idea of ML is to treat various signal processing tasks as a black box, and learn the mapping between network inputs and outputs (i.e., known information and optimization variables) through various efficient networks, including deep neural networks (DNNs) \cite{[14],[15]}, convolutional neural networks (CNNs) \cite{[16]}, recurrent neural networks (RNNs) \cite{[21]}, graph neural networks (GNNs) \cite{[17],[18],[39]}, and multi-agent reinforcement learning (MARL) \cite{[19]}. By contrast, the ``\emph{graph}" structure in GNNs can perfectly match the topology of wireless communications, enabling them to rely on topological priors to perform better with fewer training samples. Moreover, GNNs can also exploit another permutation prior to learning wireless policies more efficiently. Combining the two makes GNNs stand out from numerous networks to better address various signal processing tasks in wireless communications \cite{[18]}.

Nonetheless, since most existing GNNs belong to the ``\emph{vertex}" framework, they compress the input high-dimensional observed information into a single dimension during updating hidden representations, resulting in indistinguishable input features and serious \emph{information loss} in learning wireless policies \cite{[24],[25],[40],[41],[42],[43]}. This has triggered a rapidly growing research field, namely multidimensional graph neural networks (MDGNNs) \cite{[22],[25],[30]}, which belong to the ``\emph{edge}" framework and learn effective wireless policies by updating hidden representations of hyper-edges to reduce \emph{information loss} \cite{[26]}.
\begin{figure}[t]
\centering
    \includegraphics[scale=0.4925]{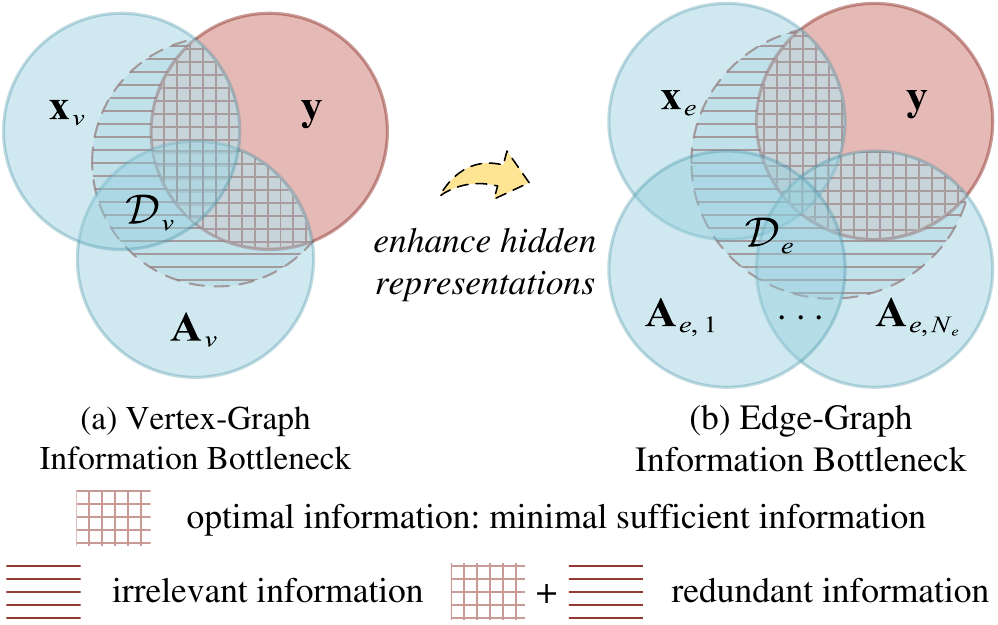}
    \caption{Illustration of GIB for signal processing, including (a) VGIB, which captures the optimal representation $\mathbf{z}_{\text{\rm v}}^l$ within input $\mathcal{D}_{\text{\rm v}}=(\mathbf{A}_{\text{\rm v}},\mathbf{x}_{\text{\rm v}})$ to predict output $\mathbf{y}$, and (b) EGIB, which captures the optimal representation $\mathbf{z}_{\text{\rm e}}^l$ within input $\mathcal{D}_\text{\rm{e}} = (\mathbf{A}_{\text{\rm{e}},1}, \mathbf{A}_{\text{\rm{e}},2}, \ldots, \mathbf{A}_{\text{\rm{e}},N_\text{\rm{e}}}, \mathbf{x}_\text{\rm{e}})$ to predict output $\mathbf{y}$.
    \label{fig1}}
\end{figure}
\vspace{-0.1cm}
\subsection{Robust Optimization Framework for Signal Processing}
Unfortunately, GNNs are often susceptible to environmental interference noise and graph structure variations during message passing, resulting in differences between the received input information and the actual observed information. Additionally, the features of adjacent nodes may contain irrelevant information that has a negative impact on the updates of the current node. To overcome these challenges and develop more powerful GNNs, we should rethink what constitutes a ``\emph{good}" hidden representation for graph-structured data. In particular, the authors in \cite{[28],[29],[37],[38]} proposed an innovative principle for representation learning, i.e., ``\emph{vertex}" information bottleneck (VIB), which suggests that the optimal hidden representation should contain the \emph{minimal sufficient} information required for optimization tasks. Correspondingly, the author in \cite{[32],[33],[35]} extended the traditional VIB to a novel ``\emph{vertex}" graph information bottleneck (VGIB) to adapt to more complex graph-structured data, overcoming the challenges posed by non-independent and identically distributed (i.i.d.) data by utilizing the local dependency assumption of graph-structured data \cite{[32]}. Specifically, both VIB and VGIB encourage the optimal hidden representation to provide maximum information about optimization variables (\emph{sufficient}) and prevent obtaining irrelevant information from input information (\emph{minimal}) \cite{[28],[29],[37],[32],[33],[35]}. This enables the learned model to effectively avoid overfitting between known input information and hidden representations, thereby better balancing the effectiveness and robustness of hidden representations in graph-structured data.

However, all existing research on IB \cite{[28],[29],[32],[33],[35],[34],[38]} focuses on ``\emph{vertex}" frameworks, whose hidden representations are updated on vertices rather than hyper-edges, resulting in indistinguishable input features and serious \emph{information loss} \cite{[24],[25],[40],[41],[42],[43]}. Therefore, effectively extract the \emph{minimal sufficient} information from hyper-edge-structured data is an important challenge that has not yet been studied in designing models.

In this paper, we strive to design robust and efficient GNNs for signal processing in wireless communications by adopting innovative representation learning principles and multiple prior knowledge. Then, we propose a robust MDGNN optimization framework with ``\emph{edge}" graph information bottleneck (EGIB), which inherits information theory principles from VGIB, as shown in Fig. 1. Specifically, EGIB utilizes hyper-edges instead of traditional vertices to enhance hidden representations while retaining the original optimal hidden representation principle, which provides maximum information about optimization variables (\emph{sufficient}) and prevents irrelevant information from being obtained from the input information (\emph{minimal}). Our main contributions are summarized as follows:
\begin{itemize}
\item We design a generic framework with EGIB for signal processing, including non-nested and nested sets. This framework effectively combines multidimensional graphs and optimal representation principles to better balance between effectiveness and robustness. Specifically, multidimensional graphs update hidden representations of hyper-edges to reduce information loss (\emph{effectiveness}), while optimal representation principles prompt representations to provide maximum information about optimization variables (\emph{effectiveness}) and prevents irrelevant information from being obtained from input information (\emph{robustness}).
\item We extend the proposed optimization framework to different structures, including permutation dimensions and objects, and analyze the differences in system performance and network overhead under various structures.
\item Extensive experimental results across various signal processing tasks in wireless communications (e.g., cell-free massive multiple-input multiple-output (mMIMO) systems), including joint precoding and power control, demonstrate the robustness and effectiveness of our proposed framework in learning wireless policies, especially in environments prone to interference.
\end{itemize}

The rest of this paper is organized as follows. Section \uppercase\expandafter{\romannumeral2} summarizes notations and a brief discussion on MDGNNs, including modeling methods and multidimensional permutation properties. The detailed GIB principle and robust MDGNN optimization framework are presented in Section \uppercase\expandafter{\romannumeral3}. The experimental evaluation demonstrating the performance of the proposed optimization framework in comparison to traditional GNNs and optimization theory-based algorithms under different signal processing tasks are shown in Section \uppercase\expandafter{\romannumeral4}. Finally, conclusions are drawn in Section \uppercase\expandafter{\romannumeral5}.

\emph{\textbf{{Notation}}}: Boldface lowercase letters $\bf{x}$ and boldface uppercase letters $\bf{X}$ denote column vectors and matrices, respectively. $\left(\cdot\right)$\textsuperscript{\emph{H}} and $\left(\cdot\right)$\textsuperscript{\emph{\textrm{T}}} represent the conjugate transpose and transpose, respectively. $\mathbb{R}^n$ and $\mathbb{C}^n$ denote the $n$-dimensional spaces of real and complex numbers, respectively.
$\mathbb{P}(\mathbf{x})$ and $\mathbb{Q}(\mathbf{x})$ are the joint probabilistic distribution function (PDF) of the random variable $\mathbf{x}$, where $\mathbb{P}(\mathbf{x})$ belongs to the induced distribution and $\mathbb{Q}(\mathbf{x})$ belongs to the variational distribution, respectively. $\mathbf{x}_1\perp\mathbf{x}_2|\mathbf{x}_3$ denotes the conditional independence of $\mathbf{x}_1$ and $\mathbf{x}_2$ given $\mathbf{x}_3$.
$\mathbb{B}(\cdot)$, $\mathbb{E}\{\cdot\}$, and $\triangleq$ are the Bernoulli distribution, expectation, and definition, respectively. $\mathcal{C}(\cdot)$ represents the hyper-edge constraint function. Finally, the circularly symmetric complex Gaussian random variable $\mathbf{x}$ with zero mean and correlation matrix $\mathbf{R}$ satisfies $\mathbf{x} \sim {{\cal N}_\mathbb{C}}(\mathbf{0},\mathbf{R})$.
\section{Multidimensional Graph Neural Networks}
In this section, we first summarize the main notions and formalize the definition of signal processing problems in wireless communications. Then, we propose an efficient MDGNN framework that updates hidden representations of hyper-edges to reduce information loss, and divide permutation problems into two categories, including non-nested and nested sets.
\vspace{-0.2cm}
\subsection{Preliminaries: Permutation and Problem Formulation}
\emph{Sets.} A set refers to a collection of objects with fixed elements but no specific order, including non-nested and nested sets, where nested sets contain independent subsets that are closer to actual communication scenarios (e.g., different types and numbers of antennas). This makes it impossible for each ``independent subset" to be permuted with each other, i.e., $\mathcal{S}_i \nleftrightarrow \mathcal{S}_j$, only allowing permutation of elements within each ``independent set", i.e., $e_{a} \leftrightarrow e_{b}$, and $e_{c} \leftrightarrow e_{d}$, $e_{a},e_{b} \in \mathcal{S}_i$, $e_{c},e_{d} \in \mathcal{S}_j$. By contrast, the type and number of elements in non-nested sets are uniform, which allows them to be permuted arbitrarily with each other, i.e., $\mathcal{S}_i \leftrightarrow \mathcal{S}_j$, $e_{a} \leftrightarrow e_{b}$, and $e_{c} \leftrightarrow e_{d}$, $e_{a},e_{b} \in \mathcal{S}_i$, $e_{c},e_{d} \in \mathcal{S}_j$.

\begin{figure}[t]
\centering
    \includegraphics[scale=0.3]{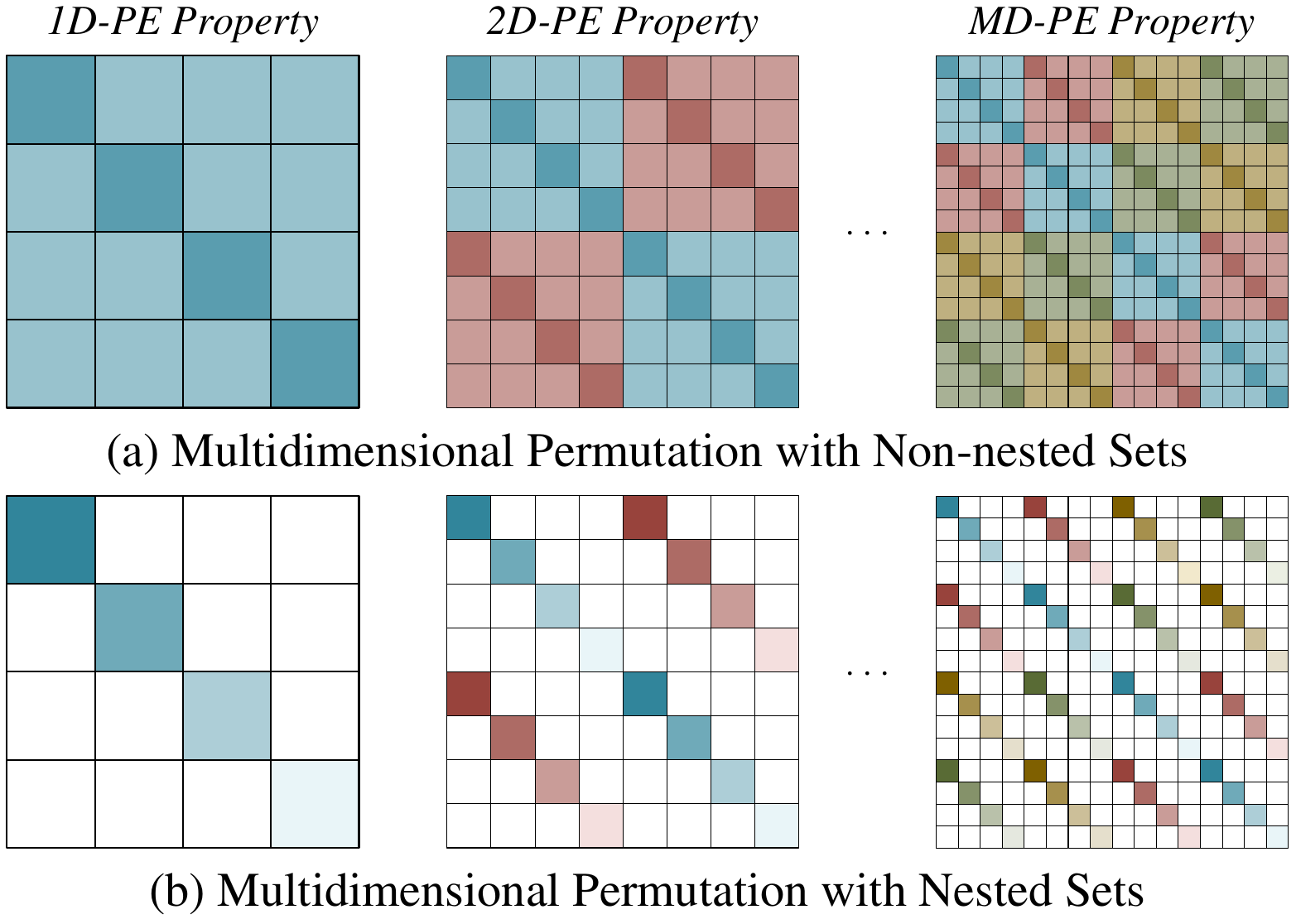}
    \caption{Structure of our multidimensional permutation matrices for the 1D-PE property (left), 2D-PE property (middle), and MD-PE property (right). Note that modules of the same color have the same coefficient, while the coefficient of the white module is zero.
    \label{fig1}}
\end{figure}
\emph{Graphs.} A graph comprises vertices, edges, and features (including vertex and edge features). In general, an edge is connected to two vertices, and an edge connected to two or more vertices can be called a hyper-edge. Moreover, the relevant features can be represented as tensors to better preserve the high-dimensional form of graph-structured data.

\emph{Problem Definition.} For a given signal processing task in wireless communications, it can usually be expressed as a multivariate optimization problem $\mathcal{P}$ with objective function $g_0(\cdot)$ and $n_c$ constraint function $g_c(\cdot)$ ($1\leqslant c\leqslant n_c$) as
\begin{subequations}
\begin{align}
\max_{\mathbf{Y}_1,\ldots,\mathbf{Y}_O} \quad&{g}_0(\mathbf{X}_1,\ldots,\mathbf{X}_I,\mathbf{Y}_1,\ldots,\mathbf{Y}_O)\\
\mathrm{s.t.}  \qquad&{g}_c(\mathbf{X}_1,\ldots,\mathbf{X}_I,\mathbf{Y}_1,\ldots,\mathbf{Y}_O)\geqslant0,
\end{align}
\end{subequations}
where $\mathbf{X}_1,\ldots,\mathbf{X}_I$ represent input environmental information, and $\mathbf{Y}_1,\ldots,\mathbf{Y}_O$ represent output optimization variables. Correspondingly, the given optimization problem can also be modeled as a multivariate signal processing policy $\mathcal{Y}=\mathcal{G}(\mathcal{X})$ with the combined input $\mathcal{X}=\{\mathbf{X}_1,\ldots,\mathbf{X}_I\}$ and combined output $\mathcal{Y}=\{\mathbf{Y}_1,\ldots,\mathbf{Y}_O\}$, and the number of input $\mathcal{X}$ and output $\mathcal{Y}$ can be different $I\neq O$.

\begin{figure}[t]
\centering
    \includegraphics[scale=0.36]{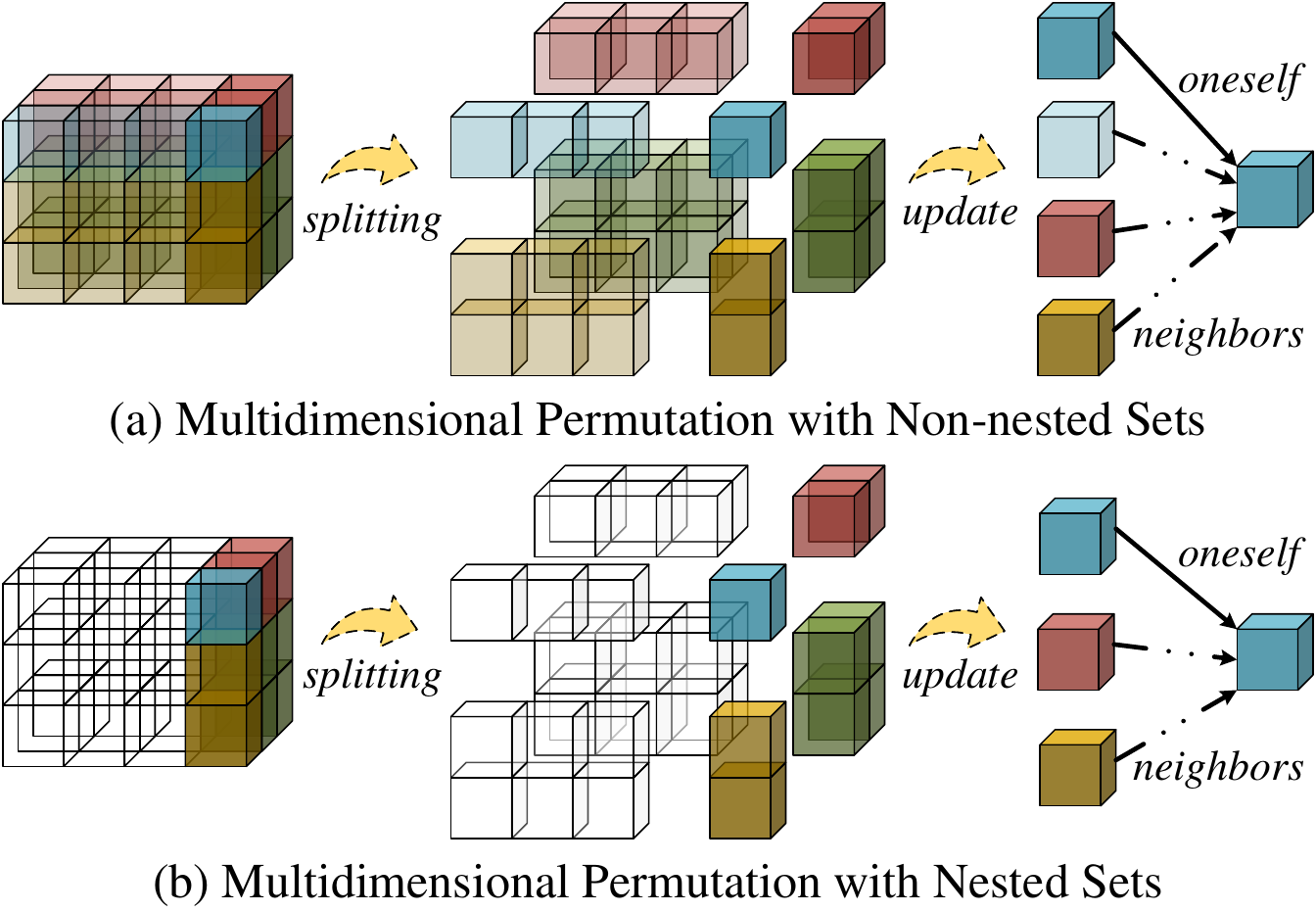}
    \caption{Update layer of MDGNNs (taking 3D as an example), including non-nested and nested sets, both of which update their own information using adjacent modules. The specific updated equation is shown in Table \uppercase\expandafter{\romannumeral2}.
    \label{fig1}}
\end{figure}
\emph{Permutation Functions.} It is a function or policy $\mathcal{Y}=\mathcal{G}(\mathcal{X})$ defined on permutation sets, where the mapping between input $\mathcal{X}$ and output $\mathcal{Y}$ remains unchanged under any permutation of a two-dimensional square matrix $\mathbf{\Pi}$, satisfying $\mathbf{\Pi}\mathcal{Y}=\mathcal{G}(\mathbf{\Pi}\mathcal{X})$. For example, when a input feature $\mathbf{X}$ is expressed as higher-order tensor, using $\pi(\cdot)$, which maps the $i$-th element of a set into the $\pi(i)$-th element, provides a more concise way to represent permutations of the tensor's elements along each dimension \cite{[25]}. Specifically, consider the edges connecting $N_1$ vertices of the first type and $N_2$ vertices of the second type (or hyper-edges connecting three or more types of vertices), permuting the input feature matrix $\mathbf{X}$ to obtain $\mathbf{\bar{X}}=\mathbf{\Pi}_1^T\mathbf{X}\mathbf{\Pi}_2$ with $(\mathbf{\bar{X}})_{i,j}=(\mathbf{X})_{\pi_1(i)\pi_2(j)}$, $i=1,\ldots,N_1$, $j=1,\ldots,N_2$, where $[\pi_k(1),\ldots,\pi_k(N_k)]^T=\mathbf{\Pi}_k^T[1,\ldots,N_k]$.

\emph{Permutation Problems.} Similarly, it is a optimization problem $\mathcal{P}$ whose the optimization objective ${g}_0$ and $n_{\text{\rm c}}$ constraints ${g}_c$ belong to permutation functions, satisfying ${g}_0(\mathbf{\Pi}\mathcal{Y},\mathbf{\Pi}\mathcal{X})={g}_0(\mathcal{Y},\mathcal{X})$ and ${g}_c(\mathbf{\Pi}\mathcal{Y},\mathbf{\Pi}\mathcal{X})={g}_c(\mathcal{Y},\mathcal{X})$, $1\leqslant c\leqslant n_c$.

Note that permutation properties of a policy belong to a type of prior knowledge, which can be utilized in multidimensional graph frameworks to improve learning efficiency. Moreover, to provide a unified framework for learning various signal processing policies with MDGNNs, we divide permutation problems $\mathcal{P}$ into two categories, including multidimensional permutation with non-nested sets $\mathcal{P}^{\text{\rm non-nested}}$ and nested sets $\mathcal{P}^{\text{\rm nested}}$. The specific permutation details are given in the following Section \uppercase\expandafter{\romannumeral2}-C and Section \uppercase\expandafter{\romannumeral2}-D, respectively.
\subsection{Method for Modeling Multidimensional Graphs}
The permutability of the problem and permutation properties of the obtained policy are induced by all sets, over which the objective ${g}_0$ and $n_{\text{\rm c}}$ constraints ${g}_c$ are defined. To model multidimensional graphs, we first find all permutation sets of optimization problems and then define corresponding hyper-edges, the specific steps are as follows:

{\emph{Step-1:}} \emph{Dimension aggregation}, aggregate all dimensions of input information $\mathcal{X}$ and output optimization variables $\mathcal{Y}$, and treat all elements in each dimension as a virtual set $\mathcal{S}_j^{\text{\rm virtual}}$, $j = 1,\ldots,J^{\text{\rm virtual}}$, where $J^{\text{\rm virtual}}$ represents the number of dimensions in the union of input $\mathcal{X}$ and output $\mathcal{Y}$.

{\emph{Step-2:}} \emph{Permutation recognition}, distinguish between non-nested sets $\mathcal{S}_a^{\text{\rm non-nested}}$ and nested sets $\mathcal{S}_b^{\text{\rm nested}}$ in all constructed virtual sets $\mathcal{S}_j^{\text{\rm virtual}}$, $a = 1,\ldots, J^{\text{\rm non-nested}}$, $b = 1,\ldots, J^{\text{\rm nested}}$, where $J^{\text{\rm non-nested}}$ and $J^{\text{\rm nested}}$ are the numbers of non-nested and nested sets, respectively. Note that subsets within nested sets are independent of each other, resulting in nested sets not having permutability.

{\emph{Step-3:}} \emph{Permutation recognition}, determine whether the optimization objective ${g}_0$ and $n_\mathrm{c}$ constraints ${g}_c$ remain unchanged after arbitrarily permutation $\mathbf{\Pi}$ of these virtual non-nested sets $\mathcal{S}_a^{\text{\rm non-nested}}$, i.e., ${g}_0(\mathbf{\Pi}\mathcal{Y},\mathbf{\Pi}\mathcal{X})={g}_0(\mathcal{Y},\mathcal{X})$ and ${g}_c(\mathbf{\Pi}\mathcal{Y},\mathbf{\Pi}\mathcal{X})={g}_c(\mathcal{Y},\mathcal{X})$, $1\leqslant c\leqslant n_c$. If so, this indicates that constructed sets has permutability and belong to a real set $\mathcal{S}_j^{\text{\rm real}}$, $j = 1,\ldots,J^{\text{\rm real}}$, where $J^{\text{\rm real}}$ represents the number of permutable sets, satisfying $J^{\text{\rm real}} \leqslant J^{\text{\rm non-nested}}$.

{\emph{Step-4:}} \emph{Vertex determination}, associate permutable sets with vertex types, and define the elements in each set of the problem $\mathcal{P}$ as vertices of each type, thereby designing an MDGNN with permutation properties that match a signal processing policy.

In this context, \emph{Step-2} and \emph{Step-3} are utilized to identify whether sets have permutability, while \emph{Step-4} is utilized to define all permutable sets as vertex types in a multidimensional framework, laying the foundation for designing MDGNNs with permutation properties for wireless communications.
\vspace{-0.3cm}
\subsection{Multidimensional Permutation with Non-nested Sets}
For a signal processing problem $\mathcal{P}$, we need to prove in advance on which sets it has permutability \cite{[25]},
which can be permutated arbitrarily without affecting the mapping between input $\mathcal{X}$ and output $\mathcal{Y}$.
Then, we can introduce permutation prior to design an efficient MDGNN framework for learning signal processing policies. Specifically, for non-nested set, as shown in Fig. 2 (a), their permutation equivariance (PE) properties in different dimensions are as follows:

\emph{One dimensional (1D)-PE Property.} A given permutation function $\mathcal{G}_1^{\text{\rm non}}$ defined on one set $\mathbf{y}_1=\mathcal{G}_1^{\text{\rm non}}(\mathbf{x}_1)$ with tensor $\mathbf{x}_1 \in \mathbb{R}^{d_1}$, satisfying, $\mathbf{\Pi}_1^T\mathbf{y}_1=\mathcal{G}_1^{\text{\rm non}}(\mathbf{\Pi}_1^T\mathbf{x}_1)$.

\emph{Two dimensional (2D)-PE Property.} A given permutation function $\mathcal{G}_2^{\text{\rm non}}$ defined on two sets $\mathbf{Y}_2=\mathcal{G}_2^{\text{\rm non}}(\mathbf{X}_2)$ with tensor $\mathbf{X}_2 \in \mathbb{R}^{d_1\times d_2}$ and vectorized version $\mathbf{x}_2 \in \mathbb{R}^{d_1d_2}$, satisfying, $\mathbf{\Pi}_1^T\mathbf{Y}\mathbf{\Pi}_2=\mathcal{G}_2^{\text{\rm non}}(\mathbf{\Pi}_1^T\mathbf{X}\mathbf{\Pi}_2)$ or $\mathbf{\Omega}_2^{\text{\rm non}}\mathbf{y}_2=\mathcal{G}_2^{\text{\rm non}}(\mathbf{\Omega}_2^{\text{\rm non}}\mathbf{x}_2)$ with $\mathbf{\Omega}_2^{\text{\rm non}}\triangleq\mathbf{\Pi}_1^T\otimes \mathbf{\Pi}_2^T$.

\emph{MD-PE Property.} A given permutation function $\mathcal{G}_J^{\text{\rm non}}$ defined on $J$ sets $\mathbf{Y}_J=\mathcal{G}_J^{\text{\rm non}}(\mathbf{X}_J)$ with tensor $\mathbf{X}_J \in \mathbb{R}^{d_1\times \cdots\times d_J}$ and  vectorized version $\mathbf{x}_J \in \mathbb{R}^{d_1d_2\cdots d_J}$, satisfying, $\mathbf{\Omega}_J^{\text{\rm non}}\mathbf{y}_J=\mathcal{G}_J^{\text{\rm non}}(\mathbf{\Omega}_J^{\text{\rm non}}\mathbf{x}_J)$ with $\mathbf{\Omega}_J^{\text{\rm non}} \triangleq \mathbf{\Pi}_1^T \otimes \mathbf{\Pi}_2^T \otimes \cdots \otimes \mathbf{\Pi}_J^T$.

\begin{rem}
To satisfy PE properties in different dimensions, the update equation of the hidden layer with the vectorized version of tensor $\mathbf{z}_j^l \in \mathbb{R}^{d_1d_2\cdots d_j}$ can be expressed as $\mathbf{z}_j^{l+1} = \sigma(\mathbf{P}^{l}\mathbf{z}_j^l) \in \mathbb{R}^{d_1d_2\cdots d_j}$ in the $l$-th layer with activation function $\sigma(\cdot)$ and $\mathbf{z}_j^1=\sigma(\mathbf{P}^{0}\mathbf{x}_j)$, $j=1.\ldots,J$, $l=1,\ldots,L-1$, where $L$ is the number of layers of MDGNNs, and $\mathbf{P}^{l}$ is the structured weight matrix to be designed. Note that we neglected the first dimension of the hidden representation composed of $N_{\text{\rm c}}$ ``channels" that are unrelated to permutations, which is from $\mathbb{R}^{N_{\text{\rm c}}d_1d_2\cdots d_j}$ to $\mathbb{R}^{d_1d_2\cdots d_j}$, thereby reducing symbol complexity.
\end{rem}
\vspace{-0.1cm}
Correspondingly, the acquisition of the structured weight matrices $\mathbf{P}^{l,{\text{\rm non-nested}}}$ with $j$ non-nested sets can adopt a method similar to that in \cite{[25]}, which has different level hierarchical parameter sharing as follows:
\begin{equation}
\setcounter{equation}{2}
\mathbf{P}_{x_{1:i}}^{l,{\text{\rm non-nested}}}=\mathbf{P}_{i,\text {\rm non-diag}}^{l} \in \mathbb{R}^{d_{i+1}\cdots d_j\times d_{i+1}\cdots d_j},
\end{equation}
where $0\leqslant i \leqslant j-1$, $x_1, x_{2}, \ldots, x_{j-1} \in \{1, 2\}$, and
\begin{equation}
\setcounter{equation}{3}
\mathbf{P}_{i,\text {\rm non-diag}}^{l}\!=\!\begin{pmatrix}
                        \mathbf{P}_{x_{1:i},1}^{l,{\text{\rm non-nested}}} & \!\!\!\mathbf{P}_{x_{1:i},2}^{l,{\text{\rm non-nested}}} & \!\!\!\cdots & \!\!\!\mathbf{P}_{x_{1:i},2}^{l,{\text{\rm non-nested}}}\\
                        \mathbf{P}_{x_{1:i},2}^{l,{\text{\rm non-nested}}} & \!\!\!\mathbf{P}_{x_{1:i},1}^{l,{\text{\rm non-nested}}} & \!\!\!\cdots & \!\!\!\mathbf{P}_{x_{1:i},2}^{l,{\text{\rm non-nested}}}\\
                        \vdots                    & \!\!\!\vdots                    & \!\!\!\ddots & \!\!\!\vdots\\
                        \mathbf{P}_{x_{1:i},2}^{l,{\text{\rm non-nested}}} & \!\!\!\mathbf{P}_{x_{1:i},2}^{l,{\text{\rm non-nested}}} & \!\!\!\cdots & \!\!\!\mathbf{P}_{x_{1:i},1}^{l,{\text{\rm non-nested}}}
                        \end{pmatrix}.
\end{equation}

Therefore, the number of trainable parameters in $\mathbf{P}^{l,{\text{\rm non-nested}}}$ decreases from $2^{d_1+d_2\ldots +d_J}$ to $2^{J}$, e.g., when $J=2$, $d_1=4$, and $d_2=4$, i.e., $2^{d_1+d_2\ldots +d_J}=256$ and $2^{J}=4$, the number of trainable parameters has been reduced by 98.44\%, effectively reducing network overhead and significantly improving learning efficiency. Moreover, considering that the number of trainable parameters $2^J$ in $\mathbf{P}^{l,{\text{\rm non-nested}}}$ still increases exponentially with the expansion of the network size, as shown in Fig. 2 (a). This prompts us to introduce topological priors to further simplify $\mathbf{P}^{l,{\text{\rm non-nested}}}$, where the hidden representation of each hyper-edge $\mathbf{z}_j^l$ only aggregates information from adjacent hyper-edges, and the weights of non-adjacent hyper-edges should be zero. This reduces the number of trainable parameters from the original $2^J$ to $J+1$, further reducing network overhead. Specifically, there is only a single dimension of variation between hyper-edge and its adjacent hyper-edges. Taking three dimensional (3D) permutation as an example, as shown in Fig. 3 (a), the update of the blue solid module is only related to its own and adjacent red solid modules, brown solid modules, and blue transparent modules, and is not associated with the other modules.
\subsection{Multidimensional Permutation with Nested Sets}
Similarly, for nested sets, as shown in Fig. 2 (b), which contain independent subsets and are different from non-nested sets, resulting in non-permutation. Therefore, their PE properties in different dimensions are as follows:

\emph{1D-PE Property.} No permutation set, i.e., $\mathbf{y_1}=\mathcal{G}_1(\mathbf{x_1})$.

\emph{2D-PE Property.} A given permutation function $\mathcal{G}_2$ defined on two sets (i.e., one pair of nested sets) $\mathbf{Y}_2=\mathcal{G}_2(\mathbf{X_2})$ with $\mathbf{X_2} \in \mathbb{R}^{d_1\times d_2}$ and $\mathbf{x_2} \in \mathbb{R}^{d_1d_2}$, satisfying, $\mathbf{\Omega}_2\mathbf{y_2}=\mathcal{G}_2(\mathbf{\Omega}_2\mathbf{x_2})$ with $\mathbf{\Omega}_2 \triangleq (\mathbf{\Pi}_1^T \otimes I_{d_2})\, \text{\rm diag} (\mathbf{\Pi}_{2,1}^T, \ldots, \mathbf{\Pi}_{2,d_1}^T)$.

\emph{MD-PE Property.} A given permutation function $\mathcal{G}_J$ defined on $J$ sets (i.e., $P$ pair of nested sets and $Q$ non-nested sets) $\mathbf{Y}_J=\mathcal{G}_J(\mathbf{X}_J)$ with $\mathbf{X}_J \in \mathbb{R}^{d_1\times d_2\times \cdots\times d_J}$ and $\mathbf{x}_J \in \mathbb{R}^{d_1 d_2 \cdots d_J}$, satisfying, $\mathbf{\Omega}_J\mathbf{y}_J=\mathcal{G}_J(\mathbf{\Omega}_J\mathbf{x}_J)$ with $\mathbf{\Omega}_{J} \triangleq \mathbf{\Omega}_{J_1} \otimes \cdots \otimes \mathbf{\Omega}_{J_P} \otimes \mathbf{\Omega}_{J_Q}$, where $2P+Q = J$, $\mathbf{\Omega}_{J_u} \triangleq (\mathbf{\Pi}_{J_{u,1}}^T \otimes I_{d_{J_{u,2}}})\, \text{\rm diag} (\mathbf{\Pi}_{{J_{u,2}},1}^T, \ldots, \mathbf{\Pi}_{{J_{u,2}},d_{J_{u,1}}}^T)$, and $\mathbf{\Omega}_{J_Q} \triangleq \mathbf{\Pi}_{J_{2P+1}}^T \otimes \cdots \otimes\mathbf{\Pi}_{J_{2P+Q}}^T$, $u=1,\ldots,P$.

Correspondingly, the structured weight matrices $\mathbf{P}^{l,{\text{\rm nested}}}$ with $p$ pair of nested and $j-2p$ non-nested sets can be represented as follows:
\begin{subequations}
\begin{align}
\mathbf{P}_{x_{1:2a-2}}^{l, \text {\rm nested}}&=\mathbf{P}_{2a-2,\text {\rm diag}}^{l} \in \mathbb{R}^{d_{2a-1}\cdots d_j\times d_{2a-1}\cdots d_j},\\
\mathbf{P}_{x_{1:2a-1}}^{l, \text {\rm nested}}&= \mathbf{P}_{2a-1,\text {\rm non-diag}}^{l} \in \mathbb{R}^{d_{2a}\cdots d_j\times d_{2a}\cdots d_j},\\
\mathbf{P}_{x_{1:b}}^{l, \text {\rm nested}}&= \mathbf{P}_{b,\text {\rm non-diag}}^{l} \in \mathbb{R}^{d_{b+1}\cdots d_{j}\times d_{b+1}\cdots d_{j}},
\end{align}
\end{subequations}
where $1\leqslant a \leqslant p$, $2p\leqslant b \leqslant j-1$, $x_{2a-2} \in \{1,\ldots,d_{2a-2}\}$, $x_{2a-1},x_{2a},x_{j-1} \in \{1,2\}$, and
\begin{equation}
\setcounter{equation}{5}
\mathbf{P}_{2a-2,\text {\rm diag}}^{l}\!=\!\begin{pmatrix}
                        \mathbf{P}_{x_{1:2a-2},1}^{l, \text {\rm nested}}  &                           &        & \\
                                                  & \!\!\!\!\mathbf{P}_{x_{1:2a-2},2}^{l, \text {\rm nested}} &        & \\
                                                  &                           & \!\!\!\!\ddots & \\
                                                  &                           &        & \!\!\!\!\mathbf{P}_{x_{1:2a-2},d_{i+1}}^{l, \text {\rm nested}}
                        \end{pmatrix}.
\end{equation}

Correspondingly, the number of trainable parameters in $\mathbf{P}^{l,{\text{\rm nested}}}$ decreases from $2^{d_1+d_2\cdots +d_J}$ to $d_1\cdots d_{2p-1}2^{p+q}$, e.g., when $J=3$, $d_1=4$, $d_2=4$, $d_3=4$, $p=1$ and $q=1$, i.e., $2^{d_1+d_2\ldots +d_J}=4096$ and $d_1\cdots d_{2p-1}2^{p+q}=16$, the number of trainable parameters has been reduced by 99.61\%, still significantly improving learning efficiency. Moreover, we also introduce topological priors to further simplify $\mathbf{P}^{l,{\text{\rm nested}}}$, reducing the number of trainable parameters from the original $d_1\cdots d_{2p-1}2^{p+q}$ to $d_1\cdots d_{2p-1}(p+q+1)$, as shown in Fig. 2 (b). Note that adjacent hyper-edges of each hyper-edge in nested sets are only related to subsets and cannot aggregate information from adjacent nested sets.

However, GNNs are susceptible to interference noise, leading to biases in graph features and structures during message passing. To address this challenge, we introduce an IB principle to learn robust hidden representations to more effectively cope with various interference noises in the following section.
\section{Robust Multidimensional Framework}
In this section, we first introduce the IB principle, which indicates that the optimal hidden representation should contain the minimal sufficient information required to avoid overfitting. Moreover, we extend it to graph-structured data, including VGIB and EGIB, to make it more robust to actual communication scenarios with high interference noise.
\vspace{-0.1cm}
\subsection{Preliminaries: Information Bottleneck}
Due to the hidden representations $\mathbf{z}^l$ (simplified version of $\mathbf{z}_j^l$) in existing MDGNNs may aggregate irrelevant information from neighbors, and MDGNNs are also susceptible to noise interference, resulting in inaccurate received information.
We introduce an innovative IB principle, which provides a critical policy for the hidden representation, i.e., the optimal representation $\mathbf{z}^l$ can contain the minimal sufficient information for signal processing problems $\mathcal{P}$, as shown in Fig. 1.

\begin{figure*}[b]
\hrulefill
\normalsize
\setcounter{equation}{12}
\begin{equation}
\begin{split}
\widehat{\text{\rm{V-VGIB}}}^l=\text{\rm{log}}\frac{\mathbb{P}(\mathbf{z}_{\text{\rm v}}^l|\mathbf{z}_{\text{\rm v}}^{l-1},\mathbf{z}_{A}^l)}{\mathbb{Q}(\mathbf{z}_{\text{\rm v}}^l)}=\sum_{v \in \mathcal{V}}\left[\text{\rm{log}}\,\Phi\left(\mathbf{z}_{\text{\rm v},v}^l;\mu_v,\sigma_v^2\right)-\text{\rm{log}}\Big(\sum_{x=1}^Xw_{x,v}\Phi\left(\mathbf{z}_{0,\text{\rm v},v}^l;\mu_{0,x},\sigma_{0,x}^2\right)\Big)\right].
\end{split}
\end{equation}
\label{eq1}
\end{figure*}
Specifically, IB encourages the hidden representation $\mathbf{z}^l$ to provide maximum information about the output $\mathbf{y}$ (\emph{sufficient}). Meanwhile, IB also prevents useless information obtained for the input $\mathbf{x}$ (\emph{minimal}). Based on this formulation, the objective of IB can be reduced to the following optimization:
\begin{equation}
\setcounter{equation}{6}
\min_{\mathbf{z}^l} \,\, \text{IB}(\mathbf{x},\mathbf{y};\mathbf{z}^l) \triangleq -I(\mathbf{y};\mathbf{z}^L) + \beta I(\mathbf{x};\mathbf{z}^L),
\end{equation}
where $\beta$ operates as a tradeoff parameter between the two terms, $\mathbf{z}^L$ represents the hidden representation of the last layer.
Moreover, $I(\mathbf{y};\mathbf{z}^L)$ represents \emph{sufficient}, while the optimization objective of conventional MDGNNs are similar, except for the introduction of $I(\mathbf{x};\mathbf{z}^L)$ related to \emph{minimal}, which helps to prevent the aggregation of redundant information and naturally avoids overfitting in the learned model. For example, for a given signal processing task, it is often impossible to obtain the corresponding wireless policies $\mathbf{y}$ in advance before training, resulting in the inability to calculate $I(\mathbf{y};\mathbf{z}^L)$. This prompts us to transform supervised objective $I(\mathbf{y};\mathbf{z}^L)$ into unsupervised loss function $\mathcal{L}(\mathbf{x})$ to overcome this challenge.
\begin{figure}[t]
\centering
    \includegraphics[scale=0.493]{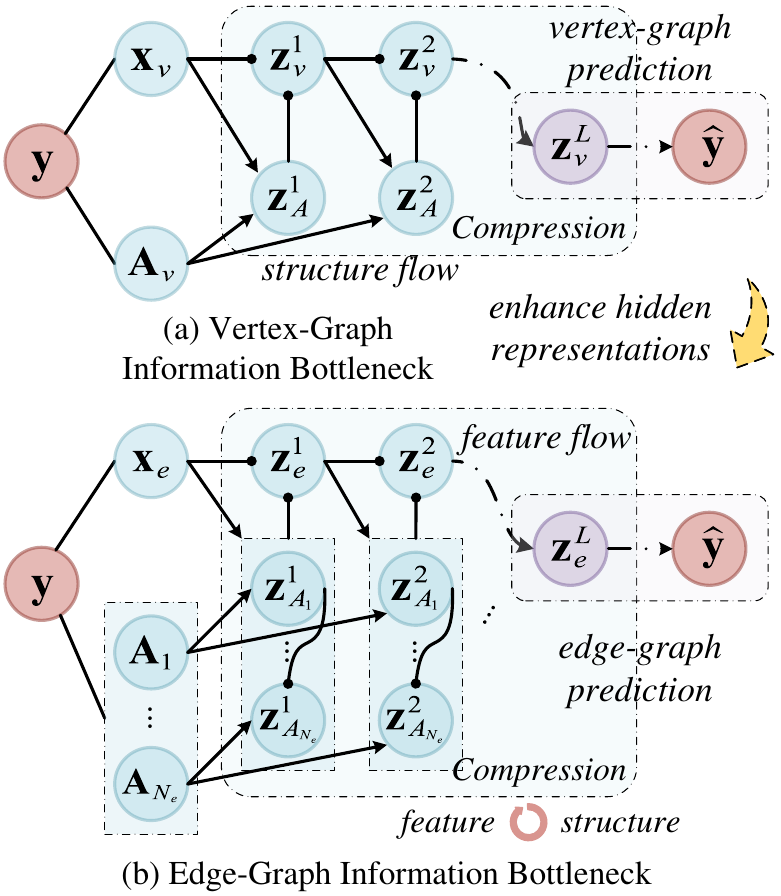}
    \caption{Markov chain of search space under GIB, including (a) VGIB, which satisfies $\mathbb{P}(\mathbf{z}_\text{\rm v}^l|\mathcal{D}_{\text{\rm v}})\in \Omega_{\text{\rm v}}$ and (b) EGIB, which satisfies $\mathbb{P}(\mathbf{z}_\text{\rm e}^l|\mathcal{D}_{\text{\rm e}})\in \Omega_{\text{\rm e}}$.
    \label{fig1}}
\end{figure}
\vspace{-0.1cm}
\subsection{Vertex-Graph Information Bottleneck}
In general, VGIB is inherited from the conventional IB principle, which requires the optimal representation $\mathbf{z}^l$ to minimize the information from the graph-structured data $\mathcal{D}_\text{\rm v}$, including the input $\mathbf{x}_\text{\rm v}$ and the newly introduced graph structure $\mathbf{A}_\text{\rm v}$, and maximize the output $\mathbf{y}$, as shown in Fig. 1 (a).

However, due to the correlation among vertices, we introduce the local-dependency assumption to constrain the space $\Omega_{\text{\rm v}}$ of the optimal representation for graph-structured data \cite{[32]}. Specifically, for a vertex $v\in \mathcal{V}$, given a certain number of hops of neighbor-related data, the rest of the data in the graph will be independent of vertex $v$, resulting in the easier to handle GIB principle, where $\mathcal{V}=\{1,\ldots,V\}$ represents the set of vertices. Therefore, we utilize ${\mathbb{P}(\mathbf{z}_\text{\rm v}^l|\mathcal{D}_{\text{\rm v}})\in \Omega_{\text{\rm v}}}$ to hierarchically iterate hidden representations to model the correlation. Then, the representation of each vertex can be refined by using a graph structure $\mathbf{z}_A^l$ to aggregate the representation of adjacent vertices. Thus, $\{\mathbf{z}_A^l\}_{1\leqslant l\leqslant L}$ can be obtained by adjusting the original graph structure $\mathbf{A}_\text{\rm v} \in \mathbb{R}^{V\times V}$, which can approximately control feature and structure flow.
Based on this formulation, the objective of VGIB can be reduced to:
\begin{equation}
\setcounter{equation}{7}
\min_{\mathbb{P}(\mathbf{z}_\text{\rm v}^l|\mathcal{D}_{\text{\rm v}})\in \Omega_{\text{\rm v}}}
\!\!\!\!\text{VGIB}(\mathcal{D}_{\text{\rm v}},\mathbf{y};\mathbf{z}_\text{\rm v}^l) \triangleq -I(\mathbf{y};\mathbf{z}_{\text{\rm v}}^L) + \beta I(\mathcal{D}_{\text{\rm v}};\mathbf{z}_{\text{\rm v}}^L),
\end{equation}
where $\Omega_{\text{\rm v}}$ is the space of the conditional distribution of $\mathbf{z}_{\text{\rm v}}^L$ given the vertex-data $\mathcal{D}_{\text{\rm v}}$ by following the probabilistic dependence shown in Fig. 4 (a). This requires us to optimize the distributions of $\mathbb{P}(\mathbf{z}_\text{\rm v}^l|\mathbf{z}_\text{\rm v}^{l-1},\mathbf{z}_A^l)$ and $\mathbb{P}(\mathbf{z}_A^l|\mathbf{z}_\text{\rm v}^{l-1},\mathbf{A}_\text{\rm v})$.

\emph{Variational Bounds:} However, the computation of $I(\mathbf{y};\mathbf{z}_{\text{\rm v}}^L)$ and $I(\mathcal{D}_{\text{\rm v}};\mathbf{z}_{\text{\rm v}}^L)$ in equation (7) is still intractable. Thus, we should introduce variational bounds on these two terms, which enables the final objective to be optimized.
As demonstrated in \cite{[32]}, we can derive the lower bound of term $I(\mathbf{y};\mathbf{z}_{\text{\rm v}}^L)$ and the upper bound of term $I(\mathcal{D}_{\text{\rm v}};\mathbf{z}_{\text{\rm v}}^L)$, which are shown in the following \textbf{Theorem 1} and \textbf{Theorem 2}, respectively.
\begin{thm}
For any variational distributions $\mathbb{Q}_1(\mathbf{y}_v|\mathbf{z}_{\text{\rm v},v}^l)$ for $v \in \mathcal{V}$ and $\mathbb{Q}_2(\mathbf{y})$, the lower bound of term $I(\mathbf{y};\mathbf{z}_{\text{\rm v}}^L)$ within VGIB can be derived as
\begin{equation}
\setcounter{equation}{8}
\begin{aligned}
I(\mathbf{y};\mathbf{z}_{\text{\rm v}}^L) \geqslant &1 + \mathbb{E}_{\mathbb{P}(\mathbf{y},\mathbf{z}_{\text{\rm v}}^L)}\left[\text{\rm{log}}\frac{\prod_{v\in \mathcal{V}}\mathbb{Q}_1(\mathbf{y}_v|\mathbf{z}_{{\text{\rm v}},v}^L)}{\mathbb{Q}_2(\mathbf{y})}\right]\\
&- \mathbb{E}_{\mathbb{P}(\mathbf{y})\mathbb{P}(\mathbf{z}_{\text{\rm v}}^L)}\left[\frac{\prod_{v\in \mathcal{V}}\mathbb{Q}_1(\mathbf{y}_v|\mathbf{z}_{{\text{\rm v}},v}^L)}{\mathbb{Q}_2(\mathbf{y})}\right].
\end{aligned}
\end{equation}
\end{thm}
\begin{IEEEproof}
The proof is given in Appendix A.
\end{IEEEproof}
\begin{thm}
For any variational distributions $\mathbb{Q}(\mathbf{z}_{A}^l)$, $l \in \mathcal{S}_A$, and $\mathbb{Q}(\mathbf{z}_{\text{\rm v}}^l)$, $l \in \mathcal{S}_{\text{\rm v}}$, the upper bound of term $I(\mathcal{D}_{\text{\rm v}};\mathbf{z}_{\text{\rm v}}^L)$ within VGIB can be derived as
\begin{equation}
\setcounter{equation}{9}
\begin{aligned}
I(\mathcal{D}_{\text{\rm v}};\mathbf{z}_{\text{\rm v}}^L) &\leqslant  I(\mathcal{D}_{\text{\rm v}};\{\mathbf{z}_{A}^l\}_{l \in \mathcal{S}_A} \cup \{\mathbf{z}_{\text{\rm v}}^l\}_{l \in \mathcal{S}_{\text{\rm v}}})\\
&\leqslant \sum_{l \in \mathcal{S}_A}\text{\rm{A-VGIB}}^l + \sum_{l \in \mathcal{S}_{\text{\rm v}}}\text{\rm{V-VGIB}}^l,
\end{aligned}
\end{equation}
where $\text{\rm{A-VGIB}}^l$ and $\text{\rm{V-VGIB}}^l$ satisfy
\begin{subequations}
\begin{align}
\text{\rm{A-VGIB}}^l&=\mathbb{E}_{\mathbb{P}(\mathbf{z}_{A}^l,\mathbf{A}_{\text{\rm v}},\mathbf{z}_{\text{\rm v}}^{l-1})}\left[\text{\rm{log}}\frac{\mathbb{P}(\mathbf{z}_{A}^l|\mathbf{A}_{\text{\rm v}},\mathbf{z}_{\text{\rm v}}^{l-1})}{\mathbb{Q}(\mathbf{z}_{A}^l)}\right],\\
\text{\rm{V-VGIB}}^l&=\mathbb{E}_{\mathbb{P}(\mathbf{z}_{\text{\rm v}}^l,\mathbf{z}_{\text{\rm v}}^{l-1},\mathbf{z}_{A}^l)}\left[\text{\rm{log}}\frac{\mathbb{P}(\mathbf{z}_{\text{\rm v}}^l|\mathbf{z}_{\text{\rm v}}^{l-1},\mathbf{z}_{A}^l)}{\mathbb{Q}(\mathbf{z}_{\text{\rm v}}^l)}\right].
\end{align}
\end{subequations}
\end{thm}
\begin{IEEEproof}
The proof is given in Appendix B.
\end{IEEEproof}
Note that index sets $\mathcal{S}_A$ and $\mathcal{S}_{\text{\rm v}}$ are introduced to ensure conditional independence between $\mathcal{D}_{\text{\rm v}}$ and $\mathbf{z}_{\text{\rm v}}^L$, which satisfies the following two properties: \emph{1) $\mathcal{S}_{\text{\rm v}} \neq \emptyset$; 2) if the maximum index  is $l$ in $\mathcal{S}_{\text{\rm v}}$, then $\mathcal{S}_A$ should contain all integers in $[l + 1, L]$.}

\emph{Objective for Training:} To optimize VGIB's parameters, we specify the bounds of term $I(\mathbf{y};\mathbf{z}_{\text{\rm v}}^L)$ and term $I(\mathcal{D}_{\text{\rm v}};\mathbf{z}_{\text{\rm v}}^L)$, and then calculate the bound of VGIB's optimization objective.

Firstly, we assume ${\mathbb{Q}(\mathbf{z}_{A}^l)}$ is a non-informative distribution and adopt a Bernoulli version as $\mathbf{z}_{A,v}^l=\cup_{t \in \mathcal{T}}\{u \in \mathcal{V}_{v,t}|u \sim \mathbb{B}(\alpha)\}$ and $\mathbf{z}_{A,v}^l \perp \mathbf{z}_{A,u}^l$ if $v \neq u$ and $\alpha \in (0,1)$. Then, we can obtain an empirical estimation of $\text{\rm{A-VGIB}}^l$ with $\phi_{v,t}^l = \sigma(\mathbf{z}_{A,v}^l,\{\mathbf{z}_{A,u}^l\}_{u\in \mathcal{V}_{v,t}})$ and constructed sets $\mathcal{V}_{v,t}$ as
\begin{equation}
\setcounter{equation}{11}
\widehat{\text{\rm{A-VGIB}}}^l=\mathbb{E}_{\mathbb{P}(\mathbf{z}_{A}^l|\mathbf{A}_{\text{\rm v}},\mathbf{z}_{\text{\rm v}}^{l-1})}\left[\text{\rm{log}}\frac{\mathbb{P}(\mathbf{z}_{A}^l|\mathbf{A}_{\text{\rm v}},\mathbf{z}_{\text{\rm v}}^{l-1})}{\mathbb{Q}(\mathbf{z}_{A}^l)}\right],
\end{equation}
which is instantiated with the Bernoulli distribution as follows:
\begin{equation}
\setcounter{equation}{12}
\widehat{\text{\rm{A-VGIB}}}^l=\sum_{v\in \mathcal{V},t\in \mathcal{T}}{\text{\rm KL}}\Big(\mathbb{B}(\phi_{v,t}^l)\parallel\mathbb{B}(\alpha)\Big).
\end{equation}

Secondly, we set $\mathbb{Q}(\mathbf{z}_{\text{\rm v}}^l)$ as a mixture of Gaussian distributions and utilize $\mathbf{z}_{{\text{\rm v}}}^l$ to estimate $\text{\rm{V-VGIB}}^l$ as (13), where $\mathbf{z}_{{\text{\rm v}},v}^l \sim \mathbb{G}(\mu_{v},\sigma_{v}^2)$ with network output $\mu_{v}$ and $\sigma_{v}^2$ based on $\mathbf{z}_{{\text{\rm v}}}^l$, and $\mathbf{z}_{0,{\text{\rm v}},v}^l \sim \sum_{x=1}^Xw_{x,v}\mathbb{G}(\mu_{0,x},\sigma_{0,x}^2)$ with shared trainable parameters $w_{x,v}$, $\mu_{0,x}$, and $\sigma_{0,x}^2$ for all vertices. Moreover, $\mathbf{z}_{{\text{\rm v}},v}^l \perp \mathbf{z}_{{\text{\rm v}},u}^l$ and $\mathbf{z}_{0,{\text{\rm v}},v}^l \perp \mathbf{z}_{0,{\text{\rm v}},u}^l$ if $v \neq u$.
Thus, we can select appropriate index sets $\mathcal{S}_{A}$ and $\mathcal{S}_{\text{\rm v}}$ that satisfy
\begin{equation}
\setcounter{equation}{14}
I(\mathcal{D}_{\text{\rm v}};\mathbf{z}_{\text{\rm v}}^L)\rightarrow \sum_{l \in \mathcal{S}_A}\widehat{\text{\rm{A-VGIB}}}^l + \sum_{l \in \mathcal{S}_{\text{\rm v}}}\widehat{\text{\rm{V-VGIB}}}^l.
\end{equation}

\begin{table*}[t]
    \caption{Features and Computational Complexity of Different Optimization Algorithms, Including Baseline and Proposed}
	\label{tab2}
	\centering
	\fontsize{8.6}{9}\selectfont
	\renewcommand{\arraystretch}{1.25}
    \begin{tabular}{lllll}
    \hline\hline
    \textbf{\makecell{Algorithms}} &\textbf{\makecell{CSI}} & \textbf{\makecell{Objects}}& \textbf{\makecell{Computational Complexity (Networks)}} &\textbf{\makecell{Optimization Objectives}} \\ \hline\hline
    \multicolumn{5}{l}{\emph{\textbf{Optimization Theory-Based}, including WMMSE and Upper Bound (WMMSE with perfect CSI)}} \\ \hline\hline
    \emph{\makecell{WMMSE}}  & \makecell{Imperfect} & \makecell{Vertices} &\makecell{Iteration - $\mathcal{O({\rm SE})}$}  & \makecell{$\max I(\mathbf{y};\mathbf{z}_\text{\rm v}^L)$} \\ \hline
    \emph{\makecell{Upper Bound}}  & \makecell{Perfect} & \makecell{Vertices} & \makecell{Iteration - $\mathcal{O({\rm SE})}$} & \makecell{$\max I(\mathbf{y};\mathbf{z}_\text{\rm v}^L)$} \\ \hline
    \multicolumn{5}{l}{\emph{\textbf{Deep Learning-Based}}, including ``regular-GNNs" (vertex-GNNs) and MDGNNs, selecting a typical type of 2D-GNN-L-K} \\ \hline\hline
    \emph{\makecell{Vertex-GNN}}  & \makecell{Imperfect} & \makecell{Vertices} & \makecell{$\mathcal{O}((L+K)(2EC+4NC+2\sum_{l=1}^{A}C^2))$} & \makecell{$\max I(\mathbf{y};\mathbf{z}_\text{\rm v}^L)$} \\
                                   &                     &                    & \makecell{$+ \mathcal{O}(LK(8NC+2\sum_{l=1}^{A}C^2))$} &  \\ \hline
    \emph{\makecell{VIB-GNN}}  & \makecell{Imperfect} & \makecell{Vertices} & \makecell{$\mathcal{O}((L+K)(2EC+4NC+2\sum_{l=1}^{A}C^2+4ES))$} & \makecell{$\max I(\mathbf{y};\mathbf{z}_\text{\rm v}^L)-\beta I(\mathbf{x}_\text{\rm v};\mathbf{z}_\text{\rm v}^L)$} \\
                               &                      &                     & \makecell{$+ \mathcal{O}(LK(8NC+2\sum_{l=1}^{A}C^2))$} &  \\ \hline
    \emph{\makecell{VGIB-Bern}}  & \makecell{Imperfect} & \makecell{Vertices} & \makecell{$\mathcal{O}((L+K)(4N_{\rm e}N+4EC+8NC+4\sum_{l=1}^{A}C^2+4ES))$} & \makecell{$\max I(\mathbf{y};\mathbf{z}_\text{\rm v}^L)-\beta I(\mathcal{D}_\text{\rm v};\mathbf{z}_\text{\rm v}^L)$} \\
                                 &                      &                     & \makecell{$+ \mathcal{O}(LK(16NC+4\sum_{l=1}^{A}C^2))$} &  \\ \hline
    \emph{\makecell{Edge-MDGNN}}  & \makecell{Imperfect} & \makecell{Edges} & \makecell{$\mathcal{O}(LK(12NC+3\sum_{l=1}^{A}C^2))$} & \makecell{$\max I(\mathbf{y};\mathbf{z}_\text{\rm e}^L)$} \\ \hline
    \textbf{\emph{\makecell{EIB-MDGNN}}}  & \makecell{Imperfect} & \makecell{Edges} & \makecell{$\mathcal{O}(LK(12NC+3\sum_{l=1}^{A}C^2+2ES))$} & \makecell{$\max I(\mathbf{y};\mathbf{z}_\text{\rm e}^L)-\beta I(\mathbf{x}_\text{\rm e};\mathbf{z}_\text{\rm e}^L)$} \\ \hline
    \textbf{\emph{\makecell{EGIB-Bern}}}  & \makecell{Imperfect} & \makecell{Edges} & \makecell{$\mathcal{O}(LK(2N_{\rm e}N+24NC+6\sum_{l=1}^{A}C^2+2ES))$} & \makecell{$\max I(\mathbf{y};\mathbf{z}_\text{\rm e}^L)-\beta I(\mathcal{D}_\text{\rm e};\mathbf{z}_\text{\rm e}^L)$} \\ \hline
\end{tabular}
\end{table*}
Finally, term $I(\mathbf{y};\mathbf{z}_{\text{\rm v}}^L)$ with the weight matrix $\mathbf{W}_\text{\rm{out}}$ reduces to cross-entropy loss (\emph{supervised}) by ignoring the constant as
\begin{equation}
\setcounter{equation}{15}
I(\mathbf{y};\mathbf{z}_{\text{\rm v}}^L)\rightarrow -\sum_{v \in \mathcal{V}}\text{\rm{Cross-Entropy}}\left(\mathbf{z}_{{\text{\rm v}},v}^L\mathbf{W}_\text{\rm{out}};\mathbf{y}_{v}\right),
\end{equation}
where $I(\mathbf{y};\mathbf{z}_{\text{\rm v}}^L)$ can be simplified as an optimization function value $\mathcal{G}(\mathbf{z}_{{\text{\rm v}},v}^L\mathbf{W}_\text{\rm{out}})$ when $\mathbf{y}_{v}$ is unknown (\emph{unsupervised}).
\begin{figure*}[b]
\hrulefill
\normalsize
\setcounter{equation}{21}
\begin{equation}
\begin{split}
\widehat{\text{\rm{E-EGIB}}}^l=\text{\rm{log}}\frac{\mathbb{P}(\mathbf{z}_\text{\rm{e}}^l|\mathbf{z}_\text{\rm{e}}^{l-1},\mathcal{Z}_{A}^e)}{\mathbb{Q}(\mathbf{z}_\text{\rm{e}}^l)}=\sum_{i \in \mathcal{N}_\text{\rm{e}}}\sum_{e_i \in \mathcal{E}_i} \left[\text{\rm{log}}\,\Phi\left(\mathbf{z}_{\text{\rm{e}},e_i}^l;\mu_{e_i},\sigma_{e_i}^2\right)-\text{\rm{log}}\Big(\sum_{x=1}^Xw_{x,e_i}\Phi\left(\mathbf{z}_{0,\text{\rm{e}},e_i}^l;\mu_{0,x},\sigma_{0,x}^2\right)\Big)\right].
\end{split}
\end{equation}
\label{eq1}
\end{figure*}
\vspace{-0.2cm}
\subsection{Edge-Graph Information Bottleneck}
However, due to the excessive compression of useful input information $\mathcal{D}_\text{\rm{v}}$ by vertices during the update of hidden representations $\mathbf{z}_\text{\rm{v}}^l$ in VGIB, the output $\mathbf{y}$ lacks accuracy. By contrast, we extend the original vertices $\mathcal{D}_\text{\rm{v}}$ to 2D edges $\mathcal{D}_\text{\rm{e},2D}$ and then to multidimensional hyper-edges $\mathcal{D}_\text{\rm{e}}$, significantly reducing information loss, where multidimensional graph-structured data $\mathcal{D}_\text{\rm{e}} = (\mathbf{A}_{\text{\rm{e}},1},\ldots, \mathbf{A}_{\text{\rm{e}},N_\text{\rm{e}}}, \mathbf{x}_\text{\rm{e}})$, edge type set $\mathcal{N}_\text{\rm{e}}=\{1,\ldots,N_\text{\rm{e}}\}$, edge set $\mathcal{E}=\cup_{i \in \mathcal{N}_\text{\rm{e}}}\{\mathcal{E}_i\}$, and 2D graph-structured data $\mathcal{D}_\text{\rm{e},2D}=\mathcal{D}_\text{\rm{e}}|_{\mathcal{N}_\text{\rm{e}}=1}$ belongs to a special case of multidimensional data, as shown in Fig. 1 (b). For example, for typical precoding optimization problems in wireless communications, channel $\mathbf{H} \in \mathbb{C}^{MKN}$ often has 3D features, including AP-set $\mathcal{M}$, UE-set $\mathcal{K}$, and antenna-set $\mathcal{N}$, and directly training it in the network seriously loses important channel features. This prompts us to expand $\mathbf{H} \in \mathbb{C}^{MKN}$ into 2D channel $\mathbf{H}_{\text{\rm{2D}}} \in \mathbb{C}^{MK \times N}$ or 3D channel $\mathbf{H}_{\text{\rm{3D}}} \in \mathbb{C}^{M\times K \times N}$ to preserve important channel features and suppress information loss. Moreover, the number of edge types $N_e$ and the corresponding graph structures $\mathcal{A}_e$ can be determined based on the 3D channel $\mathbf{H}_{\text{\rm{3D}}} \in \mathbb{C}^{M\times K \times N}$, e.g., based on the existing vertex types, including AP-set $\mathcal{M}$, UE-set $\mathcal{K}$, and antenna-set $\mathcal{N}$. Similarly, we utilize ${\mathbb{P}(\mathbf{z}_\text{\rm{e}}^l|\mathcal{D}_\text{\rm{e}})\in \Omega_\text{\rm{e}}}$ to hierarchically iterate representations, and then the representation of each hyper-edge can be refined by using the graph structure $\mathbf{z}_{A_{i}}^l$ to aggregate the representation of adjacent hyper-edges, where $\{\mathbf{z}_{A_{i}}^{l}\}_{1 \leqslant l \leqslant L, i \in \mathcal{N}_\text{\rm{e}}}$ can be obtained by adjusting the original graph structure $\mathcal{A}_\text{\rm{e}}=\{\mathbf{A}_{\text{\rm{e}},1},\ldots,\mathbf{A}_{\text{\rm{e}},N_\text{\rm{e}}}\}$. Based on this formulation, the objective of EGIB can be reduced to:
\begin{equation}
\setcounter{equation}{16}
\min_{\mathbb{P}(\mathbf{z}_\text{\rm{e}}^l|\mathcal{D}_\text{\rm{e}})\in \Omega_\text{\rm{e}}} \!\!\!\text{EGIB}(\mathcal{D}_\text{\rm{e}},\mathbf{y};\mathbf{z}_\text{\rm{e}}^l) \triangleq -I(\mathbf{y};\mathbf{z}_\text{\rm{e}}^L) + \beta I(\mathcal{D}_\text{\rm{e}};\mathbf{z}_\text{\rm{e}}^L),
\end{equation}
where $\Omega_\text{\rm{e}}$ is the space of the conditional distribution of $\mathbf{z}_\text{\rm{e}}^L$ given the edge-data $\mathcal{D}_\text{\rm{e}}$ by following the probabilistic dependence shown in Fig. 4 (b). Moreover, we also introduce variational bounds to derive the lower bound of term $I(\mathbf{y};\mathbf{z}_\text{\rm{e}}^L)$ and the upper bound of term $I(\mathcal{D}_\text{\rm{e}};\mathbf{z}_\text{\rm{e}}^L)$, which are shown in the
following \textbf{Corollary 1} and \textbf{Corollary 2}, respectively.

\begin{coro}
For any variational distributions $\mathbb{Q}_1(\mathbf{y}_{e_i}|\mathbf{z}_{\text{\rm{e}},{e_i}}^L)$ for $i \in \mathcal{N}_\text{\rm{e}}$, ${e_i} \in \mathcal{E}_i$, and $\mathbb{Q}_2(\mathbf{y})$, the lower bound of term $I(\mathbf{y};\mathbf{z}_\text{\rm{e}}^L)$ within EGIB can be derived as
\begin{equation}
\setcounter{equation}{17}
\begin{aligned}
\!\!\!\!I(\mathbf{y};\mathbf{z}_\text{\rm{e}}^L) \!&\geqslant\! 1 \!+\! \mathbb{E}_{\mathbb{P}(\mathbf{y},\mathbf{z}_\text{\rm{e}}^L)}\!\bigg[\text{\rm{log}}\frac{\prod_{i\in \mathcal{N}_\text{\rm{e}}}\prod_{e_i\in \mathcal{E}_i}\!\mathbb{Q}_1(\mathbf{y}_{e_i}|\mathbf{z}_{\text{\rm{e}},{e_i}}^L)}{\mathbb{Q}_2(\mathbf{y})}\bigg]\\
&-\! \mathbb{E}_{\mathbb{P}(\mathbf{y})\mathbb{P}(\mathbf{z}_\text{\rm{e}}^L)}\!\bigg[\frac{\prod_{i\in \mathcal{N}_\text{\rm{e}}}\prod_{{e_i}\in \mathcal{E}_i}\!\mathbb{Q}_1(\mathbf{y}_{e_i}|\mathbf{z}_{\text{\rm{e}},{e_i}}^L)}{\mathbb{Q}_2(\mathbf{y})}\bigg].
\end{aligned}
\end{equation}
\end{coro}
\begin{IEEEproof}
The proof is given in Appendix C.
\end{IEEEproof}
\begin{algorithm}[t]
\label{algo:AIRMN}
\caption{Robust EGIB for Signal Processing}
    \KwIn{The input environment data $\mathcal{D}_{\text{\rm e}}=(\mathcal{A}_{\text{\rm e}},\mathbf{x}_{\text{\rm e}})$;\\
     The integral limitation to impose local dependence $\mathcal{T}_i$.}
    {\bf Initiation:} $\mathbf{z}_{\text{\rm e}}^0 \leftarrow \mathbf{x}_{\text{\rm e}}$; Sets $\mathcal{E}_{e_i,t} \leftarrow \{e_j \in \mathcal{E}|\mathcal{C}(e_i,e_j)=t\}$; Weights $\mathbf{W}^1 \in \mathbb{R}^{f\times 2f'}$, $\mathbf{W}^l \in \mathbb{R}^{f'\times 2f'}$, $\mathbf{W}_{\text{\rm out}}^{e_i} \in \mathbb{R}^{f'\times d_i}$, where $\mathbf{W}^1$ and $\mathbf{W}^l$ correspond to the structured weight matrices $\mathbf{P}^{l,\text{\rm non-nested}}$ or $\mathbf{P}^{l,\text{\rm nested}}$ in MDGNNs. \\
    \KwOut{Variables $\mathbf{y}_{e_i} \leftarrow \text{\rm softmax}(\mathbf{z}_{\text{\rm e},e_i}^L\mathbf{W}_{\text{\rm out}}^{e_i})$.}
    \For{hidden layers $l \geqslant 1$ and edges $e_i \in \mathcal{E}_i$, $i\in \mathcal{N}_{\text{\rm e}}$}
        {
        $\tilde{\mathbf{z}}_{\text{\rm e},e_i}^{l-1}\leftarrow \tau(\mathbf{z}_{\text{\rm e},e_i}^{l-1})\mathbf{W}^l$\\
        \For{$t \in \mathcal{T}_i$ with constructed sets $\mathcal{E}_{e_i,t}$}
        {
        $\phi_{e_i,t}^{l} = \text{\rm softmax}(\mathbf{z}_{\text{\rm{e}},e_i}^{l},\{\mathbf{z}_{\text{\rm{e}},e_j}^{l}\}_{e_j\in \mathcal{E}_{e_i,t}})$\\
        $\mathbf{z}_{A_i,e_i}^{l}\leftarrow\cup_{t_i \in \mathcal{T}}\{e_j \in \mathcal{E}_{e_i,t}|e_i \overset{\text{\rm iid}}{\sim} \mathbb{B}(\phi_{e_i,t}^{l})\}$\\
        }
        $\bar{\mathbf{z}}_{\text{\rm e},e_i}^{l}\leftarrow \sum_{e_j \in \mathbf{z}_{A_i,e_i}^{l}}\tilde{\mathbf{z}}_{\text{\rm e},e_i}^{l-1}$\\
        mean $\mu_{e_i,l}\leftarrow \bar{\mathbf{z}}_{\text{\rm e},e_i}^{l}[0:f']$\\
        variance $\sigma_{e_i,l}^2\leftarrow \text{\rm softplus}(\bar{\mathbf{z}}_{\text{\rm e},e_i}^{l}[f':2f'])$\\
        $\mathbf{z}_{{\text{\rm e}},e_i}^l \sim \text{\rm Gaussian}(\mu_{e_i,l},\sigma_{e_i,l}^{2})$
        }
\end{algorithm}
\begin{table*}[t]
  \centering
  \fontsize{8.6}{9}\selectfont
  \caption{Robust EGIB for Signal Processing Optimization Problems under Different Structures in Cell-Free mMIMO Systems}
  \vspace{-2mm}
  \label{CF1}
   \begin{tabular}{                                                     !{\vrule width1.2pt}  m{1.8 cm}<{\centering}                         !{\vrule width1.2pt}  m{3.6 cm}<{\centering}                         !{\vrule width1.2pt} m{1.8cm}<{\centering}                          !{\vrule width1.2pt} m{1.7 cm}<{\centering}                          !{\vrule width1.2pt} m{6.8 cm}<{\centering}   !{\vrule width1.2pt} }

    \Xhline{1.2pt}
        \rowcolor{gray!30} \bf Classification &  \bf Permutable Policy &  \bf Training Mechanism &  \bf Trainable Parameter & \bf State Update Equation \cr
    \Xhline{1.2pt}
        \multirow{11}{*}{ \shortstack{\bf Non-nested \\\bf Sets}} & 1D-GNN: AP-set & Centralized & $2C_lC_{l+1}$ & $\mathbf{x}_{m}^{l+1}=\mathbf{P}_1^l\mathbf{x}_{m}^l + \mathbf{P}_2^l\sum_{i=1,i\neq m}^M\mathbf{x}_{i}^l$ \\

        \cline{2-5} & 1D-GNN: UE-set & Distributed  & $2C_lC_{l+1}$ & $\mathbf{x}_{k}^{l+1}=\mathbf{P}_1^l\mathbf{x}_{k}^l + \mathbf{P}_2^l\sum_{j=1,j\neq k}^K \mathbf{x}_{j}^l$\\

        \cline{2-5} & 1D-GNN: Antenna-set & Distributed & $2C_lC_{l+1}$& $\mathbf{x}_{n}^{l+1}=\mathbf{P}_1^l\mathbf{x}_{n}^l + \mathbf{P}_2^l\sum_{q=1,q\neq n}^N\mathbf{x}_{q}^l$ \\

        \cline{2-5} & 2D-GNN: AP-set and UE-set & Centralized & $3C_lC_{l+1}$ & $\mathbf{x}_{m,k}^{l+1}=\mathbf{P}_{1,1}^l\mathbf{x}_{m,k}^l + \mathbf{P}_{1,2}^l\sum_{j=1,j\neq k}^K\mathbf{x}_{m,j}^l + \mathbf{P}_{2,1}^l\sum_{i=1,i\neq m}^M\mathbf{x}_{i,k}^l$ \\

        \cline{2-5} & 2D-GNN: AP-set and Antenna-set & Centralized & $3C_lC_{l+1}$ & $\mathbf{x}_{m,n}^{l+1}=\mathbf{P}_{1,1}^l\mathbf{x}_{m,n}^l + \mathbf{P}_{1,2}^l\sum_{q=1,j\neq n}^N\mathbf{x}_{m,q}^l +  \mathbf{P}_{2,1}^l\sum_{i=1,i\neq m}^M\mathbf{x}_{i,n}^l$ \\

        \cline{2-5} & 2D-GNN: UE-set and Antenna-set & Distributed & $3C_lC_{l+1}$ & $\mathbf{x}_{k,n}^{l+1}=\mathbf{P}_{1,1}^l\mathbf{x}_{k,n}^l + \mathbf{P}_{1,2}^l\sum_{q=1,q\neq n}^N\mathbf{x}_{k,q}^l +  \mathbf{P}_{2,1}^l\sum_{j=1,j\neq k}^K\mathbf{x}_{j,n}^l$ \\

        \cline{2-5} & 3D-GNN: AP-set, UE-set and Antenna-set & Centralized  & $4C_lC_{l+1}$ & $\mathbf{x}_{m,k,n}^{l+1}=\mathbf{P}_{1,1,1}^l\mathbf{x}_{m,k,n}^l+ \mathbf{P}_{1,1,2}^l\sum_{q=1,q\neq n}^N\mathbf{x}_{m,k,q}^l+ \mathbf{P}_{1,2,1}^l\sum_{j=1,j\neq k}^K\mathbf{x}_{m,j,n}^l + \mathbf{P}_{2,1,1}^l\sum_{i=1,i\neq m}^M\mathbf{x}_{i,k,n}^l$ \cr\Xhline{1.2pt}

        \multirow{4}{*}{ \shortstack{\bf Nested Sets}} & 1D-GNN: Non-set & Distributed & $MC_lC_{l+1}$  & $\mathbf{x}_{m,n}^{l+1}=\mathbf{P}_{1}^{l,m}\mathbf{x}_{m,n}^l $ \\

        \cline{2-5} & 2D-GNN: UE-set & Distributed & $2MC_lC_{l+1}$  & $\mathbf{x}_{m,k}^{l+1}=\mathbf{P}_{1}^{l,m}\mathbf{x}_{m,k}^l + \mathbf{P}_{2}^{l,m}\sum_{j=1,j\neq k}^K\mathbf{x}_{m,j}^l$ \\

        \cline{2-5} & 2D-GNN: Antenna-set & Distributed & $2MC_lC_{l+1}$  & $\mathbf{x}_{m,n}^{l+1}=\mathbf{P}_{1}^{l,m}\mathbf{x}_{m,n}^l  + \mathbf{P}_{2}^{l,m}\sum_{q=1,q\neq n}^N\mathbf{x}_{m,q}^l$ \\

        \cline{2-5} & 3D-GNN: UE-set and Antenna-set & Distributed & $3MC_lC_{l+1}$ & $\mathbf{x}_{m,k,n}^{l+1}= \mathbf{P}_{1,1}^{l,m}\mathbf{x}_{m,k,n}^l + \mathbf{P}_{1,2}^{l,m}\sum_{q=1,q\neq n}^N\mathbf{x}_{m,k,q}^l + \mathbf{P}_{2,1}^{l,m}\sum_{j=1,j\neq k}^K\mathbf{x}_{m,j,n}^l$  \cr\Xhline{1.2pt}
    \end{tabular}
  \vspace{0cm}
\end{table*}
\begin{coro}
For any variational distributions $\mathbb{Q}(\mathbf{z}_{A_i}^{l})$, $i \in \mathcal{N}_\text{\rm{e}}$, $l \in \mathcal{S}_{A_i}$, and $\mathbb{Q}(\mathbf{z}_\text{\rm{e}}^l)$, $l \in \mathcal{S}_\text{\rm{e}}$, the upper bound of term $I(\mathcal{D}_\text{\rm{e}};\mathbf{z}_\text{\rm{e}}^L)$ within EGIB can be derived as
\begin{equation}
\setcounter{equation}{18}
\begin{aligned}
I(\mathcal{D}_\text{\rm{e}};\mathbf{z}_\text{\rm{e}}^L) &\leqslant  I(\mathcal{D}_\text{\rm{e}};\{\cup_{i \in \mathcal{N}_\text{\rm{e}}}\{\mathbf{z}_{A_i}^{l}\}_{l \in \mathcal{S}_{A_i}}\} \cup \{\mathbf{z}_\text{\rm{e}}^l\}_{l \in \mathcal{S}_\text{\rm{e}}})\\
&\leqslant \sum_{i \in \mathcal{N}_\text{\rm{e}}}\sum_{l \in \mathcal{S}_{A_i}}\text{\rm{A-EGIB}}_{i}^{l} + \sum_{l \in \mathcal{S}_\text{\rm{e}}}\text{\rm{E-EGIB}}^l,
\end{aligned}
\end{equation}
where $\text{\rm{A-EGIB}}_1^{l}$, $\text{\rm{A-EGIB}}_i^{l}$ with $\mathcal{Z}_{i}^{\text{\rm e}}=\{\mathbf{z}_{A_1}^{l},\ldots,\mathbf{z}_{A_{i-1}}^{l}\}$, $i \geqslant 2$, and $\text{\rm{E-EGIB}}^l$ with $\mathcal{Z}_{A}^{\text{\rm e}}=\{\mathbf{z}_{A_1}^l,\ldots,\mathbf{z}_{A_{{N}_\text{\rm{e}}}}^l\}$ satisfy
\begin{subequations}
\begin{align}
\text{\rm{A-EGIB}}_{1}^{l}&= \mathbb{E}_{\mathbb{P}(\mathbf{z}_{A_1}^{l},\mathbf{A}_{\text{\rm{e},1}},\mathbf{z}_{\text{\rm{e}}}^{l-1})}\bigg[\text{\rm{log}}\frac{\mathbb{P}(\mathbf{z}_{A_1}^{l}|\mathbf{A}_{\text{\rm{e},1}},\mathbf{z}_{\text{\rm{e}}}^{l-1})}{\mathbb{Q}(\mathbf{z}_{A_1}^{l})}\bigg],\\
\text{\rm{A-EGIB}}_{i}^{l}&= \mathbb{E}_{\mathbb{P}(\mathbf{z}_{A_i}^{l},\mathbf{A}_{\text{\rm{e},i}},\mathbf{z}_{\text{\rm{e}}}^{l-1},\mathcal{Z}_{i}^{\text{\rm e}})}\bigg[\text{\rm{log}}\frac{\mathbb{P}(\mathbf{z}_{A_i}^{l}|\mathbf{A}_{\text{\rm{e},i}},\mathbf{z}_{\text{\rm{e}}}^{l-1},\mathcal{Z}_{i}^{\text{\rm e}})}{\mathbb{Q}(\mathbf{z}_{A_i}^{l})}\bigg],\\ \text{\rm{E-EGIB}}^l&=\mathbb{E}_{\mathbb{P}(\mathbf{z}_\text{\rm{e}}^l,\mathbf{z}_\text{\rm{e}}^{l-1},\mathcal{Z}_{A}^{\text{\rm e}})}\bigg[\text{\rm{log}}\frac{\mathbb{P}(\mathbf{z}_\text{\rm{e}}^l|\mathbf{z}_\text{\rm{e}}^{l-1},\mathcal{Z}_{A}^{\text{\rm e}})}{\mathbb{Q}(\mathbf{z}_\text{\rm{e}}^l)}\bigg].
\end{align}
\end{subequations}
\end{coro}
\begin{IEEEproof}
The proof is given in Appendix D.
\end{IEEEproof}

Moreover, the bounds of term $I(\mathbf{y};\mathbf{z}_\text{\rm{e}}^L)$ and term $I(\mathcal{D}_\text{\rm{e}};\mathbf{z}_\text{\rm{e}}^L)$ can be similarly specified as follows:
\begin{subequations}
\begin{align}
&I(\mathcal{D}_\text{\rm{e}};\mathbf{z}_\text{\rm{e}}^L)\!\rightarrow \!\sum_{i \in \mathcal{N}_\text{\rm{e}}}\sum_{l \in \mathcal{S}_{A_i}}\!\widehat{\text{\rm{A-EGIB}}}_i^{l} + \sum_{l \in \mathcal{S}_\text{\rm{e}}}\widehat{\text{\rm{E-EGIB}}}^l\!,\\
&I(\mathbf{y};\mathbf{z}_\text{\rm{e}}^L)\!\rightarrow\! -\!\sum_{i \in \mathcal{N}_\text{\rm{e}}}\sum_{e_i \in \mathcal{E}_i}\!\text{\rm{Cross-Entropy}}\!\left(\mathbf{z}_{\text{\rm{e}},e_i}^L\!\mathbf{W}_\text{\rm{out}}^{e_i};\mathbf{y}_{e_i}\right)\!,
\end{align}
\end{subequations}
where $I(\mathcal{D}_\text{\rm{e}};\mathbf{z}_\text{\rm{e}}^L)$ can be simplified as an optimization function value $\mathcal{G}(\mathbf{z}_{\text{\rm{e}},e_i}^L\mathbf{W}_\text{\rm{out}}^{e_i})$ when $\mathbf{y}_{e_i}$ is unknown, $\widehat{\text{\rm{A-EGIB}}}_{i}^{l}$ satisfies
\begin{equation}
\setcounter{equation}{21}
\widehat{\text{\rm{A-EGIB}}}_{i}^{l}=\!\!\sum_{e_i\in \mathcal{E}_i,t\in \mathcal{T}_i}\text{\rm{KL}}\Big(\mathbb{B}(\phi_{e_i,t}^{l})\parallel\mathbb{B}(\alpha)\Big), i \in \mathcal{N}_\text{\rm{e}},
\end{equation}
and $\widehat{\text{\rm{E-EGIB}}}^l$ is shown as (22).
Based on the aforementioned study, Algorithm 1 outlines the training process of the proposed EGIB. Note that $\mathbf{z}_{0,{\text{\rm e}},e_i}^l \sim \sum_{x=1}^Xw_{x,e_i}{\text{\rm{Gaussian}}}(\mu_{0,x},\sigma_{0,x}^2)$ with shared parameters $w_{x,e_i}$, $\mu_{0,x}$, and $\sigma_{0,x}^2$.
\section{Experimental Analysis}
This section empirically evaluates the proposed EGIB focusing on the following research questions:
\begin{itemize}
\item \emph{Q1.} How does EGIB perform in enhancing the effectiveness of signal processing tasks? (\emph{Case 1 and Case 2})
\item \emph{Q2.} How does EGIB perform in improving robustness in environments prone to interference? (\emph{Case 1 and Case 2})
\item \emph{Q3.} How do multidimensional graphs and GIB principles influence the performance of EGIB? (\emph{Case 1})
\item \emph{Q4.} Can EGIB be trained stably and robustly in different interference environments? (\emph{Case 1})
\end{itemize}
\subsection{Baseline Models}
In this work, we consider various state-of-the-art baselines, including WMMSE \cite{[8],[9]} as an example of traditional optimization theory-based, and vertex-GNN \cite{[19],[24]}, VIB-GNN \cite{[28],[29]}, VGIB with Bernoulli (VGIB-Bern) \cite{[32],[33]}, and edge-MDGNN \cite{[22],[25],[30]}, as shown in Table \uppercase\expandafter{\romannumeral1}. Moreover, we select multiple frameworks with different permutation dimensions and objects for comparison.
\subsection{Case 1: Robust Framework for Optimizing Joint Precoding}
\emph{Experimental Setup:} To evaluate the performance of our proposed frameworks (including ``edge" information bottleneck (EIB)-MDGNN and EGIB with Bernoulli (EGIB-Bern)) on the joint precoding optimization problem, we conduct a series of experiments in a classic cell-free mMIMO system consisting of $M$ APs and $K$ single-antenna UEs, where all APs with $N$ antennas are connected to one central processing unit (CPU) via fronthaul links, and are assumed to be randomly distributed in a square area of $\text{1} \times \text{1}$ $\text{km}^2$ with a wrap-around scheme \cite{[2]}. In the data transmission phase, all APs serve all UEs exploiting the same time-frequency resources, then the received data signal at UE $k$ can be denoted as ${y}_k = \sum_{m=1}^{M}\sum_{i=1}^{K}\mathbf{h}_{mk}^H\mathbf{w}_{mi}s_i + n_k$, where $\mathbf{h}_{mk} \in \mathbb{C}^{N \times 1}$ denotes the channel vector between AP $m$ and UE $k$, in which the classical Rayleigh fading channel is adopted, and $\mathbf{w}_{mi} \in \mathbb{C}^{N \times 1}$ represents the precoding vector which should satisfy the power constraint of AP $m$ as $\sum_{i=1}^K\mathrm{tr}(\mathbf{w}_{mi}\mathbf{w}_{mi}^H) \leqslant p_m$. Additionally, $s_i \sim {{\cal N}_\mathbb{C}}\left({0},1\right)$ and $n_k \sim {{\cal N}_\mathbb{C}}\left({0},\sigma^2\right)$ are the data symbol and the independent receiver noise, respectively.
\begin{figure*}[t]
\centering
    \vspace{-0.3cm}
    \includegraphics[scale=0.411]{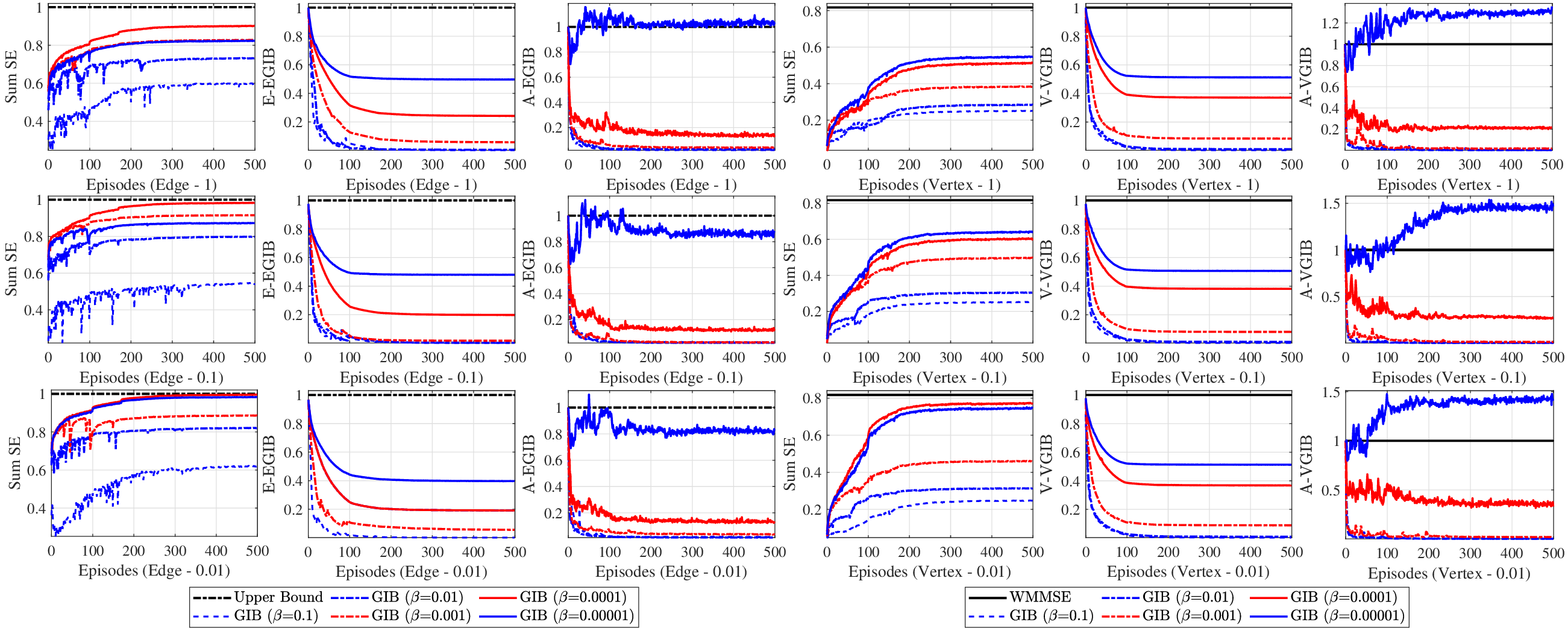}
    \vspace{-0.4cm}
    \caption{Sum SE and mutual information (including $\text{\rm E-EGIB}$, $\text{\rm A-EGIB}$, $\text{\rm V-VGIB}$, and $\text{\rm A-VGIB}$) against interference noise $\sigma_\mathrm{i}^2$ and GIB parameters $\beta$ over different optimization frameworks with $M=10$, $K=4$, and $N=4$, including EGIB and VGIB (Case 1).
    \label{fig1}}
\end{figure*}

\setcounter{equation}{22}
Then, all precoding vectors of APs in cell-free mMIMO systems can be jointly optimized, say from the following weighted sum SE maximization problem that takes into account of user fairness with $\mathbf{E}_{m,ki}=|\mathbf{h}_{mk}^H\mathbf{w}_{mi}|^2$ as
\begin{subequations}
\begin{align}
\max_{\mathbf{w}_{mk}} \sum_{k=1}^{K}\alpha_k{\text{\rm log}}_{2}&\bigg(1+\frac{\sum_{m=1}^M\mathbf{E}_{m,kk}}{\sum_{i=1,i\neq k}^K\sum_{m=1}^M\mathbf{E}_{m,ki} + \sigma^2}\bigg)\\
\mathrm{s.t.}    \thinspace\thinspace\thinspace\thinspace\qquad \qquad & \sum_{i=1}^K\mathrm{tr}(\mathbf{w}_{mi}\mathbf{w}_{mi}^H) \leqslant p_m, \forall m, i,
\end{align}
\end{subequations}
where $\alpha_k \in [0,1]$ is the fairness weight among UEs, (23a) is the precoding optimization objective, and (23b) is the power constraint.
Moreover, the pathloss model adopted to construct the channel model $\mathbf{h}_{mk}$ can be calculated by the COST 321 Walﬁsh-Ikegami model \cite{[3]}. And we consider $\sigma^2 = -94$ dBm noise power, $B = 20$ MHz communication bandwidth, and $p_m = 1$ W maximum transmit power per AP.

\emph{Permutation Properties:} Regarding the joint precoding optimization problem $\mathbf{w}_3=\mathcal{G}_3(\mathbf{h}_3)$ with vectorized variables $\mathbf{w}_3$ and $\mathbf{h}_3 \in \mathbb{C}^{MKN}$, as proven in \cite{[25]}, which belongs to a permutation problem of AP-set $\mathcal{M}=\{1,\ldots,M\}$, UE-set $\mathcal{K}=\{1,\ldots,K\}$, antenna-set $\mathcal{N}=\{1,\ldots,N\}$ in non-nested sets, and belongs to a permutation problem of UE-set $\mathcal{K}$ and antenna-set $\mathcal{N}$ in nested sets, with AP-set $\mathcal{M}$ being non-permutation, as shown in Table \uppercase\expandafter{\romannumeral2}. With the increase of permutation dimensions and objects, the hidden state update equation gradually expands to multiple dimensions, which can effectively reduce information loss, but this brings about an increase in trainable parameters and network overhead. Therefore, choosing appropriate permutation dimensions and objects is beneficial for achieving a balance between system performance and network overhead. Specifically, the elements in each set can be permuted arbitrarily, and the constructed sets can be permuted independently, satisfying 3D-PE property as
\begin{subequations}
\begin{align}
(\mathbf{\Pi}_1^T \otimes \mathbf{\Pi}_2^T \otimes \mathbf{\Pi}_3^T)\mathbf{w}_3&=\mathcal{G}_3((\mathbf{\Pi}_1^T \otimes \mathbf{\Pi}_2^T \otimes \mathbf{\Pi}_3^T)\mathbf{h}_3),\\
(\mathbf{\Omega}_2\otimes \mathbf{\Pi}_{3}^T)\mathbf{w}_3&=\mathcal{G}_3((\mathbf{\Omega}_2\otimes \mathbf{\Pi}_{3}^T)\mathbf{h}_3),
\end{align}
\end{subequations}
where 3D-non-nested permutation $\mathbf{\Omega}_3^{\text{\rm non}} \triangleq \mathbf{\Pi}_1^T \otimes \mathbf{\Pi}_2^T \otimes \mathbf{\Pi}_3^T$, and 3D-nested permutation $\mathbf{\Omega}_3 \triangleq \mathbf{\Omega}_2\otimes \mathbf{\Pi}_{3}^T$ with 2D-nested permutation $\mathbf{\Omega}_2 \triangleq (\mathbf{\Pi}_1^T \otimes I_{N}) \text{\rm diag} (\mathbf{\Pi}_{2,1}^T, \ldots, \mathbf{\Pi}_{2,M}^T)$. Note that 3D-PE property may require $4C_lC_{l+1}$ (``non-nested sets") or $3MC_lC_{l+1}$ (``nested sets") trainable parameters with the number of ``channels" $C_l$ in the $l$-th layer, which have a high network overhead. For simplify, we introduce 1D-PE property and 2D-PE property with lower trainable parameters as
\begin{subequations}
\begin{align}
\mathbf{\Pi}_1^T\mathbf{w}_1&=\mathcal{G}_1(\mathbf{\Pi}_1^T\mathbf{h}_1),\\
\mathbf{\Omega}_1\mathbf{w}_1&=\mathcal{G}_1(\mathbf{\Omega}_1\mathbf{h}_1),\\
(\mathbf{\Pi}_1^T \otimes \mathbf{\Pi}_2^T)\mathbf{w}_2&=\mathcal{G}_2((\mathbf{\Pi}_1^T \otimes \mathbf{\Pi}_2^T)\mathbf{h}_2),\\
\mathbf{\Omega}_2\mathbf{w}_2&=\mathcal{G}_2(\mathbf{\Omega}_2\mathbf{h}_2),
\end{align}
\end{subequations}
where 1D-non-nested permutation $\mathbf{\Omega}_1^{\text{\rm non}} \triangleq \mathbf{\Pi}_1^T$ for any single set with $\mathbf{w}_1,\mathbf{h}_1 \in \mathbb{C}^{d_1 \times d_2d_3}$, 2D-non-nested permutation $\mathbf{\Omega}_2^{\text{\rm non}} \triangleq \mathbf{\Pi}_1^T \otimes \mathbf{\Pi}_2^T$ for any joint set composed of two sets with $\mathbf{w}_2,\mathbf{h}_2 \in \mathbb{C}^{d_1d_2 \times d_3}$, and 1D-nested permutation $\mathbf{\Omega}_1=\text{\rm diag} (\mathbf{\Pi}_{1}^T, \ldots, \mathbf{\Pi}_{M}^T)$ for non-set. Furthermore, under GIB, we can transform optimization problem (23) based on the graph-structured data $\mathcal{D}_{\text{\rm v}}=(\mathbf{A}_{\text{\rm v}},\mathbf{h})$ and $\mathcal{D}_{\text{\rm e}}=(\mathbf{A}_{\text{\rm e},1},\ldots,\mathbf{A}_{\text{\rm e},{N}_{\text{\rm e}}},\mathbf{h})$ with different graph structures into
\begin{subequations}
\begin{align}
\!\!\!\!\min_{\mathbb{P}(\mathbf{z}_\text{\rm v}^l|\mathcal{D}_{\text{\rm v}})\in \Omega_{\text{\rm v}}}
\!\!\!\!\!\text{VGIB}(\mathcal{D}_{\text{\rm v}},\mathbf{w};\mathbf{z}_\text{\rm v}^l) \!&=\! -I(\mathbf{w};\mathbf{z}_{\text{\rm v}}^L) + \beta I(\mathcal{D}_{\text{\rm v}};\mathbf{z}_{\text{\rm v}}^L),\\
\!\!\!\!\min_{\mathbb{P}(\mathbf{z}_\text{\rm e}^l|\mathcal{D}_{\text{\rm e}})\in \Omega_{\text{\rm e}}}
\!\!\!\!\!\text{EGIB}(\mathcal{D}_{\text{\rm e}},\mathbf{w};\mathbf{z}_\text{\rm e}^l) \!&=\! -I(\mathbf{w};\mathbf{z}_{\text{\rm e}}^L) + \beta I(\mathcal{D}_{\text{\rm e}};\mathbf{z}_{\text{\rm e}}^L).
\end{align}
\end{subequations}

\begin{table*}[t]
	\centering
	\caption{Sum SE against Interference Noise $\sigma_{\mathrm{i}}^2$ under Different Structures with $M=10$, $K=4$, and $N=4$ (Case 1)}\label{table:OFDM_M}
    \fontsize{8.6}{6.9}\selectfont
\label{table:defense_adverserial}
\resizebox{\textwidth}{!}{
\begin{tabular}{l|c|ccc|ccc|c}
\hline\hline
  & \multicolumn{1}{c|}{\multirow{2}{*}{\textbf{Framework}}} & \multicolumn{7}{c}{\textbf{Sum-rate relative to WMMSE: 24.75 (bps/Hz) with $\sigma_{\mathrm{i}}^2=0.1$ and 17.54 (bps/Hz) with $\sigma_{\mathrm{i}}^2=1$}}                                                          \\ \cline{3-9}
& \multicolumn{1}{c|}{}  & \textbf{1D-GNN-L}   & \textbf{1D-GNN-K}             & \textbf{1D-GNN-U}             & \textbf{2D-GNN-L-K}             & \textbf{2D-GNN-L-U}
& \textbf{2D-GNN-K-U}             & \textbf{3D-GNN-L-K-U}                       \\ \hline
\parbox[t]{2mm}{\multirow{6}{*}{\rotatebox[origin=c]{90}{\textbf{Non-nested (0.1)}}}}

                                   & Vertex-GNN          & 16.13\scriptsize{(-34.83\%)}     & 16.13\scriptsize{(-34.83\%)} & 16.13\scriptsize{(-34.83\%)} & 16.13\scriptsize{(-34.83\%)}& 16.13\scriptsize{(-34.83\%)}      & 16.13\scriptsize{(-34.83\%)}     & 16.13\scriptsize{(-34.83\%)}\\
                                   & Edge-MDGNN          & 25.37\scriptsize{(+2.48\%)}     &  21.35\scriptsize{(-13.77\%)}     &  19.31\scriptsize{(-22.01\%)}      &  27.25\scriptsize{(+10.09\%)} &  25.88\scriptsize{(+4.57\%)}      &  24.74\scriptsize{(-0.01\%)}      & 25.19\scriptsize{(+1.78\%)}\\
                                   & VIB-GNN             & 18.23\scriptsize{(-26.34\%)}     &  18.23\scriptsize{(-26.34\%)} & 18.23\scriptsize{(-26.34\%)} & 18.23\scriptsize{(-26.34\%)} & 18.23\scriptsize{(-26.34\%)} & 18.23\scriptsize{(-26.34\%)} & 18.23\scriptsize{(-26.34\%)} \\
                                   & VGIB-Bern           & 7.44\scriptsize{(-69.94\%)}     &  7.44\scriptsize{(-69.94\%)}     &  7.44\scriptsize{(-69.94\%)}     & 7.44\scriptsize{(-69.94\%)} & 7.44\scriptsize{(-69.94\%)} & 7.44\scriptsize{(-69.94\%)} & 7.44\scriptsize{(-69.94\%)} \\ \cline{2-9}
                                   & \textbf{EIB-MDGNN}  & \underline{26.65\scriptsize{(\textbf{+7.67\%})}}    &  \underline{22.43\scriptsize{(\textbf{-9.37\%})}}     &  \underline{19.71\scriptsize{(\textbf{-20.35\%})}}     &  \underline{29.74\scriptsize{(\textbf{+20.16\%})}} &  \underline{28.91\scriptsize{(\textbf{+16.81\%})}}      & \underline{28.07\scriptsize{(\textbf{+13.41\%})}}     & \underline{29.16\scriptsize{(\textbf{+17.82\%})}} \\
                                   & \textbf{EGIB-Bern}  & \textbf{26.93}\scriptsize{(\textbf{+8.81\%})}    &  \textbf{22.75}\scriptsize{(\textbf{-8.08\%})}     &  \textbf{20.08}\scriptsize{(\textbf{-18.87\%})}     &  \textbf{30.04}\scriptsize{(\textbf{+21.37\%})} &  \textbf{29.21}\scriptsize{(\textbf{+18.02\%})}      & \textbf{28.35}\scriptsize{(\textbf{+14.55\%})}     & \textbf{29.57}\scriptsize{(\textbf{+19.47\%})} \\
                                   \hline\hline
\parbox[t]{2mm}{\multirow{6}{*}{\rotatebox[origin=c]{90}{\textbf{Nested (0.1)}}}}
                                   & Vertex-GNN          & 16.13\scriptsize{(-34.83\%)}     & 16.13\scriptsize{(-34.83\%)} & 16.13\scriptsize{(-34.83\%)} & 16.13\scriptsize{(-34.83\%)}& 16.13\scriptsize{(-34.83\%)}      & 16.13\scriptsize{(-34.83\%)}     & 16.13\scriptsize{(-34.83\%)}\\
                                   & Edge-MDGNN          & 24.64\scriptsize{(-0.41\%)}     &  21.35\scriptsize{(-13.77\%)}     &  19.31\scriptsize{(-22.01\%)}     & 26.73\scriptsize{(+8.01\%)} &  25.43\scriptsize{(+2.75\%)}      &  24.74\scriptsize{(-0.01\%)}      & 24.71\scriptsize{(-0.16\%)}\\
                                   & VIB-GNN             & 18.23\scriptsize{(-26.34\%)}     &  18.23\scriptsize{(-26.34\%)} & 18.23\scriptsize{(-26.34\%)} & 18.23\scriptsize{(-26.34\%)} & 18.23\scriptsize{(-26.34\%)} & 18.23\scriptsize{(-26.34\%)} & 18.23\scriptsize{(-26.34\%)} \\
                                   & VGIB-Bern           & 7.44\scriptsize{(-69.94\%)}     &  7.44\scriptsize{(-69.94\%)}     &  7.44\scriptsize{(-69.94\%)}     & 7.44\scriptsize{(-69.94\%)} & 7.44\scriptsize{(-69.94\%)} & 7.44\scriptsize{(-69.94\%)} & 7.44\scriptsize{(-69.94\%)} \\ \cline{2-9}
                                   & \textbf{EIB-MDGNN}  & \underline{26.14\scriptsize{(\textbf{+5.51\%})}}     &  \underline{22.43\scriptsize{(\textbf{-9.37\%})}}     &  \underline{19.71\scriptsize{(\textbf{-20.35\%})}}     &  \underline{29.37\scriptsize{(\textbf{+18.67\%})}} &  \underline{28.53\scriptsize{(\textbf{+15.27\%})}}      &  \underline{28.07\scriptsize{(\textbf{+13.41\%})}}      & \underline{28.75\scriptsize{(\textbf{+16.16\%})}}\\
                                   & \textbf{EGIB-Bern}  & \textbf{26.45}\scriptsize{(\textbf{+6.87\%})}    &  \textbf{22.75}\scriptsize{(\textbf{-8.08\%})}     &  \textbf{20.08}\scriptsize{(\textbf{-18.87\%})}     &  \textbf{29.69}\scriptsize{(\textbf{+19.96\%})} &  \textbf{28.94}\scriptsize{(\textbf{+16.93\%})}      & \textbf{28.35}\scriptsize{(\textbf{+14.55\%})}     & \textbf{29.19}\scriptsize{(\textbf{+17.94\%})} \\
                                   \hline\hline
\parbox[t]{2mm}{\multirow{6}{*}{\rotatebox[origin=c]{90}{\textbf{Non-nested (1)}}}}
                                   & Vertex-GNN          & 8.71\scriptsize{(-50.31\%)}  &  8.71\scriptsize{(-50.31\%)} &  8.71\scriptsize{(-50.31\%)} &  8.71\scriptsize{(-50.31\%)} &  8.71\scriptsize{(-50.31\%)} &  8.71\scriptsize{(-50.31\%)} &  8.71\scriptsize{(-50.31\%)} \\
                                   & Edge-MDGNN          & 20.79\scriptsize{(+18.59\%)} & 17.01\scriptsize{(-3.01\%)} & 12.28\scriptsize{(-29.96\%)} &  22.63\scriptsize{(+29.05\%)}& 20.49\scriptsize{(+16.84\%)} & 19.44\scriptsize{(+10.85\%)} & 21.41\scriptsize{(+22.07\%)} \\
                                   & VIB-GNN             & 15.51\scriptsize{(-11.53\%)}     & 15.51\scriptsize{(-11.53\%)}    & 15.51\scriptsize{(-11.53\%)} &15.51\scriptsize{(-11.53\%)} & 15.51\scriptsize{(-11.53\%)} & 15.51\scriptsize{(-11.53\%)} & 15.51\scriptsize{(-11.53\%)} \\
                                   & VGIB-Bern           & 7.16\scriptsize{(-59.18\%)}     & 7.16\scriptsize{(-59.18\%)}  & 7.16\scriptsize{(-59.18\%)}  & 7.16\scriptsize{(-59.18\%)} & 7.16\scriptsize{(-59.18\%)}  & 7.16\scriptsize{(-59.18\%)} & 7.16\scriptsize{(-59.18\%)}  \\ \cline{2-9}
                                   & \textbf{EIB-MDGNN}  & \underline{25.51\scriptsize{(\textbf{+45.46\%})}}     &  \underline{21.42\scriptsize{(\textbf{+22.17\%})}}     &  \underline{18.04\scriptsize{(\textbf{+2.87\%})}}     &  \underline{27.27\scriptsize{(\textbf{+55.51\%})}} &  \underline{25.58\scriptsize{(\textbf{+45.89\%})}} & \underline{24.51\scriptsize{(\textbf{+39.75\%})}}       &  \underline{26.17\scriptsize{(\textbf{+49.21\%})}}       \\
                                   & \textbf{EGIB-Bern}  & \textbf{25.82}\scriptsize{(\textbf{+47.21\%})}     &  \textbf{21.69}\scriptsize{(\textbf{+23.66\%})}     &  \textbf{18.32}\scriptsize{(\textbf{+4.45\%})}     &  \textbf{27.59}\scriptsize{(\textbf{+57.31\%})} &  \textbf{25.89}\scriptsize{(\textbf{+47.61\%})} & \textbf{24.77}\scriptsize{(\textbf{+41.22\%})}       &  \textbf{26.51}\scriptsize{(\textbf{+51.14\%})}       \\
                                   \hline\hline
\parbox[t]{2mm}{\multirow{6}{*}{\rotatebox[origin=c]{90}{\textbf{Nested (1)}}}}
                                   & Vertex-GNN          & 8.71\scriptsize{(-50.31\%)}  &  8.71\scriptsize{(-50.31\%)} &  8.71\scriptsize{(-50.31\%)} &  8.71\scriptsize{(-50.31\%)} &  8.71\scriptsize{(-50.31\%)} &  8.71\scriptsize{(-50.31\%)} &  8.71\scriptsize{(-50.31\%)} \\
                                   & Edge-MDGNN          & 18.84\scriptsize{(+7.42\%)} & 17.01\scriptsize{(-3.01\%)} & 12.28\scriptsize{(-29.96\%)} & 21.36\scriptsize{(+21.81\%)} &  19.61\scriptsize{(+11.82\%)}      &  19.44\scriptsize{(+10.85\%)} & 20.31\scriptsize{(+15.84\%)}\\
                                   & VIB-GNN             & 15.51\scriptsize{(-11.53\%)}     & 15.51\scriptsize{(-11.53\%)}    & 15.51\scriptsize{(-11.53\%)} &15.51\scriptsize{(-11.53\%)} & 15.51\scriptsize{(-11.53\%)} & 15.51\scriptsize{(-11.53\%)} & 15.51\scriptsize{(-11.53\%)} \\
                                   & VGIB-Bern           & 7.16\scriptsize{(-59.18\%)}     & 7.16\scriptsize{(-59.18\%)}  & 7.16\scriptsize{(-59.18\%)}  & 7.16\scriptsize{(-59.18\%)} & 7.16\scriptsize{(-59.18\%)}  & 7.16\scriptsize{(-59.18\%)} & 7.16\scriptsize{(-59.18\%)}  \\ \cline{2-9}
                                   & \textbf{EIB-MDGNN}  & \underline{24.48\scriptsize{(\textbf{+39.61\%})}} & \underline{21.42\scriptsize{(\textbf{+22.17\%})}} & \underline{18.04\scriptsize{(\textbf{+2.87\%})}} &  \underline{26.61\scriptsize{(\textbf{+51.71\%})}} &  \underline{24.93\scriptsize{(\textbf{+42.15\%})}}     &  \underline{24.51\scriptsize{(\textbf{+39.75\%})}}  & \underline{25.41\scriptsize{(\textbf{+44.87\%})}} \\
                                   & \textbf{EGIB-Bern}  & \textbf{24.85}\scriptsize{(\textbf{+41.68\%})}     &  \textbf{21.69}\scriptsize{(\textbf{+23.66\%})}     &  \textbf{18.32}\scriptsize{(\textbf{+4.45\%})}     &  \textbf{26.92}\scriptsize{(\textbf{+53.48\%})} &  \textbf{25.23}\scriptsize{(\textbf{+43.84\%})} & \textbf{24.77}\scriptsize{(\textbf{+41.22\%})}       &  \textbf{25.69}\scriptsize{(\textbf{+46.47\%})}       \\
                                   \hline\hline
\end{tabular}
}
\end{table*}
Moreover, considering the influence of uncertain factors such as multipath interference and noise in actual complex dynamic environments,
the information observed by each agent is different from the perfect channel state information (CSI) and belongs to imperfect CSI. To ensure that the designed framework is close to actual environments, we introduce random interference noise $\sigma_{\mathrm{i}}$ in the simulation settings (which increases from $10^{-2}$ to $10^{1}$ corresponding to a signal-to-interference plus noise ratio of $20$dB to $-10$dB) to verify the robustness and effectiveness of the proposed optimization framework. Meanwhile, we also adopt the perfect CSI-based WMMSE scheme as the performance upper bound baseline for subsequent comparisons, which is assumed to have no noise interference compared to other baselines compared and the proposed optimization frameworks.

\emph{Graph Information Bottleneck Parameters:} The comparison of GIB-based optimization frameworks for different parameters $\beta$ are shown in Fig. 5, including interference noise $\sigma_{\text{\rm i}}=1$, $\sigma_{\text{\rm i}}=0.1$, and $\sigma_{\text{\rm i}}=10^{-2}$. We can observe that as parameter $\beta$ decreases, the sum SE and mutual information show approximately the same growth trend under different interference noises (except for the excessively small $\beta=10^{-5}$), gradually approaching the upper bound (\emph{edge}) or WMMSE (\emph{vertex}). To enhance the performance of the proposed framework for signal processing, we select parameter $\beta=10^{-4}$ with the best sum SE performance for subsequent experiments, which can approach the performance upper bound with an error tolerance of nearly 72.23\% (E-EGIB) and 87.29\% (A-EGIB) for data transmission to better cope with random interference noise.

\emph{Results and Analysis:} The joint precoding results are presented in Table \uppercase\expandafter{\romannumeral1} (\emph{overhead}) and Table \uppercase\expandafter{\romannumeral3} (\emph{performance}). The proposed EIB-MDGNN and EGIB-Bern achieve superior performance with approximate computational computational compared to the five baselines under different structures, including permutation dimensions and objects. This indicates that introducing an IB or GIB can effectively avoid overfitting of hidden representations (achieving the goal of anti-interference), thereby achieving a better balance between performance and overhead. Meanwhile, the extended GIB enables EGIB-Bern to achieve superior sum SE performance compared to EIB-MDGNN due to further extraction of the minimal sufficient information from graph-structured data, indicating that it is more suitable for signal processing tasks with topological structures. Compared with 1D frameworks for permutation of single sets, 2D and 3D frameworks for permutation of joint sets can reduce information loss by updating the hidden representations of hyper-edges, significantly improving system performance. This demonstrates the necessity of using multidimensional frameworks that update hidden representations with edges or hyper-edges in learning wireless policies. Moreover, the system performance under architectures containing permutation objects AP-set are significantly superior to other architectures due to their centralized mechanism, which can better utilize the collaboration among APs to optimize precoding vectors. For nested sets, compared to non-nested sets, due to the independence of antenna-set under each AP, additional trainable parameters are required to optimize precoding vectors, which not only increases network overhead but also results in an inevitable loss of system performance. In the following results and analysis, considering system performance and network overhead indicators, we select the superior 2D-GNN-L-K as the basic structure for subsequent comparisons.

Next, we investigate the effectiveness and robustness of the proposed optimization framework in environments prone to interference. Fig. 6 shows the sum SE performance against the interference noise $\sigma_{\text{\rm i}}^2$ under different optimization frameworks with $M=10$, $K=4$, and $N=4$. As observed, compared to traditional baselines, including WMMSE and different GNN frameworks, the proposed EIB-MDGNN and EGIB-Bern still achieve superior performance and can better cope with random interference noise in a practical environment. More importantly, it is interesting to find that as the interference noise $\sigma_{\text{\rm i}}$ gradually increases, the performance gap between the proposed optimization frameworks and traditional baselines gradually widens, e.g., the performance gap increases from 5.74\% ($\sigma_{\text{\rm i}}^2=10^{-2}$) to 62.19\% ($\sigma_{\text{\rm i}}^2=10^{0.5}$) compared with WMMSE. This indicates that our proposed frameworks have superior robustness in environments prone to interference. Correspondingly, as the interference noise $\sigma_{\text{\rm i}}$ gradually decreases or approaches zero interference, EIB-MDGNN and EGIB-Bern can gradually approach the performance upper bound, indicating that our proposed frameworks can also achieve a better balance between effectiveness and robustness.

\begin{figure}[t]
\centering
    \includegraphics[scale=0.476]{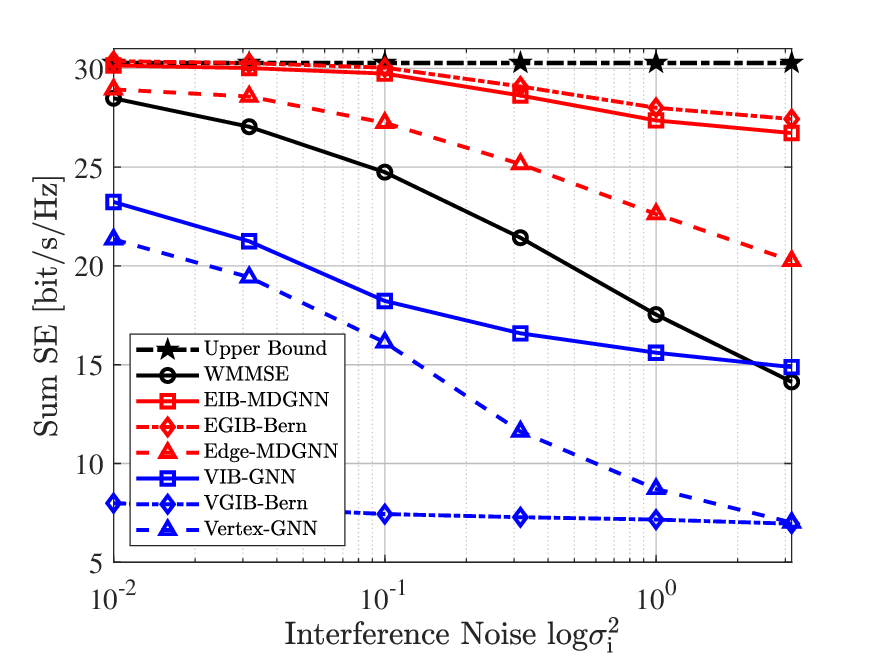}
    \caption{Sum SE against interference noise $\sigma_\mathrm{i}^2$ over different frameworks with $M=10$, $K=4$, and $N=4$ (Case 1).}
\end{figure}
\begin{figure}[t]
\centering
    \includegraphics[scale=0.476]{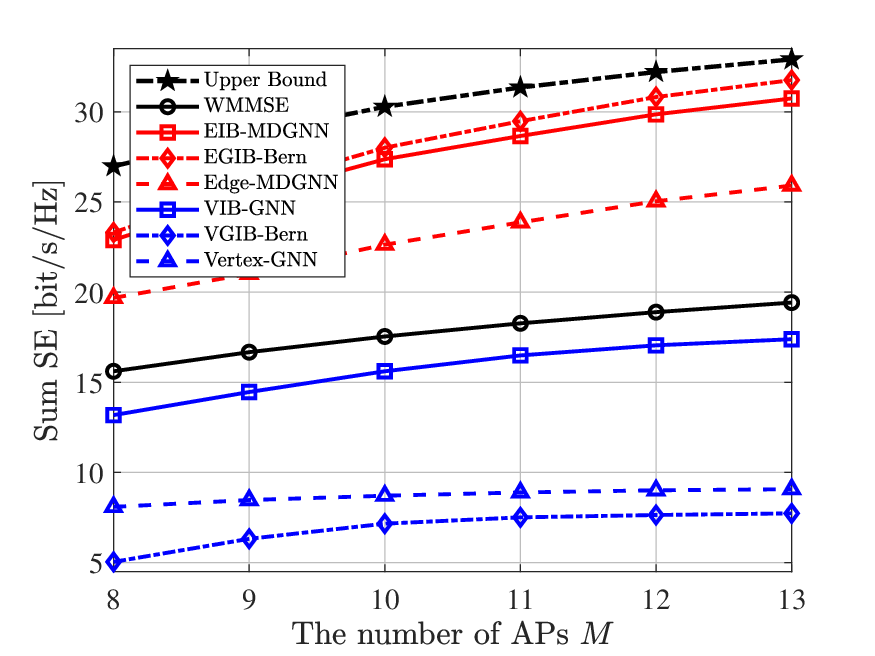}
    \caption{Sum SE against the number of APs $M$ over different frameworks with $K=4$, $N=4$, and $\sigma_\mathrm{i}^2=0.1$ (Case 1).}
\end{figure}
\begin{figure}[t]
\centering
    \includegraphics[scale=0.476]{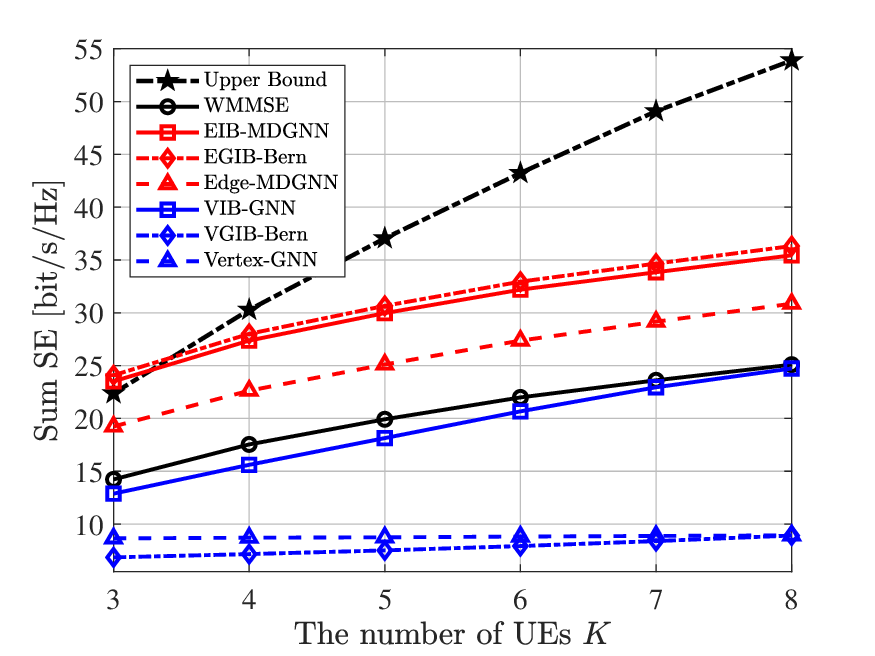}
    \caption{Sum SE against the number of UEs $K$ over different frameworks with $M=10$, $N=4$, and $\sigma_\mathrm{i}^2=0.1$ (Case 1).}
\end{figure}
\begin{figure}[t]
\centering
    \includegraphics[scale=0.476]{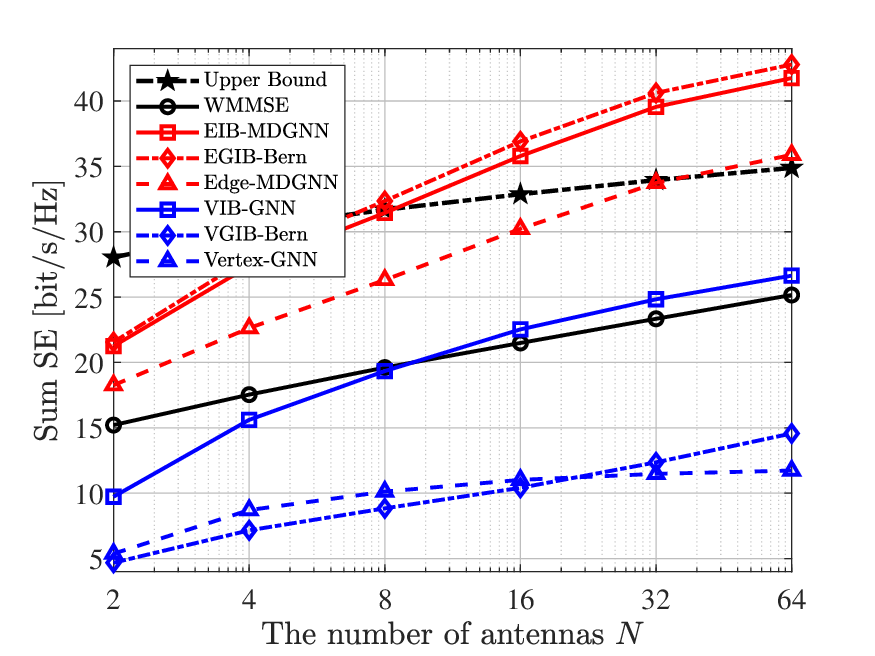}
    \caption{Sum SE against the number of antennas per AP $N$ over different frameworks with $M=10$, $K=4$, and $\sigma_\mathrm{i}^2=0.1$ (Case 1).}
\end{figure}
Moreover, to verify that the proposed optimization framework can be robust to different environments with superior performance, we compare the proposed frameworks with other baselines under different simulation parameters, including different numbers of APs ($M \in \{8,9,10,11,12,13\}$), UEs ($K \in \{3,4,5,6,7,8\}$), and antennas ($N \in \{2,4,8,16,32,64\}$), as shown in Fig. 7, Fig. 8, and Fig. 9, respectively. We can observe that the sum SE performance under the proposed EIB-MDGNN and EGIB-Bern consistently outperforms traditional WMMSE and GNN frameworks, indicating that it can adapt well to changes in various system parameters, especially in the case of a large number of antennas. For example, as the number of antennas increases from $2$ to $64$, our proposed EIB-MDGNN and EGIB-Bern gradually outperform WMMSE significantly, with the performance gap increasing from 39.71\% ($N=2$) to 69.45\% ($N=64$). More importantly, the proposed EIB-MDGNN and EGIB-Bern even exceed the performance upper bound (perfect CSI) under a large number of antennas ($N > 8$), revealing their strong robustness in environments prone to interference. This also indicates that effectively suppressing interference between antennas is more conducive to breaking through the performance upper bound.

Additionally, to further validate the generalization of our proposed frameworks, we divided the ``transferability" results into two conditions for comparative analysis, including condition 1: same configurations for training and testing and condition 2: different configurations for training and testing (e.g., train on a small number of UEs, i.e., $K=3$ and test on a much larger number of UEs, i.e., $4\leqslant K \leqslant 8$). As shown in Fig. 10 and Fig. 11, we observed a decrease in performance for all frameworks under condition 2 compared to condition 1. By contrast, the decrease in the proposed EIB-MDGNN and EGIB-Bern is much smaller than that of traditional GNN frameworks. This indicates that our proposed frameworks have good generalization. Meanwhile, compared to ``vertex" frameworks, innovative ``edge" frameworks can effectively avoid performance degradation in unknown environmental configurations by capturing channel features in different dimensions. This further highlights the importance of utilizing ``edge" frameworks in wireless communications.
\subsection{Case 2: Robust Framework for Optimizing Power Control}
\begin{figure}[t]
\centering
    \includegraphics[scale=0.45]{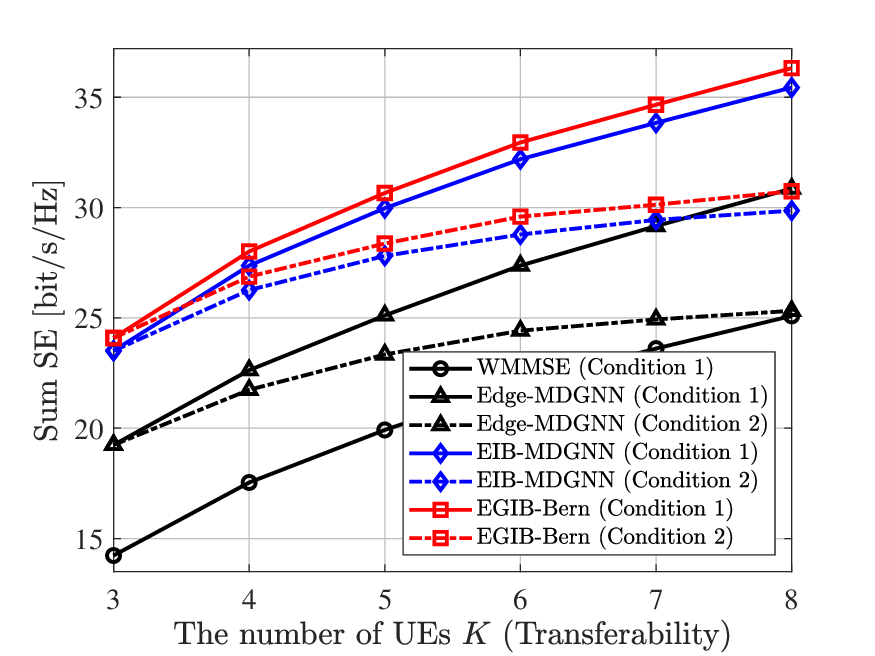}
    \caption{Sum SE against the number of UEs $K$ over different frameworks with $M=10$, $N=4$, and $\sigma_\mathrm{i}^2=0.1$ (Edge-Transferability).}
\end{figure}
\begin{figure}[t]
\centering
    \includegraphics[scale=0.45]{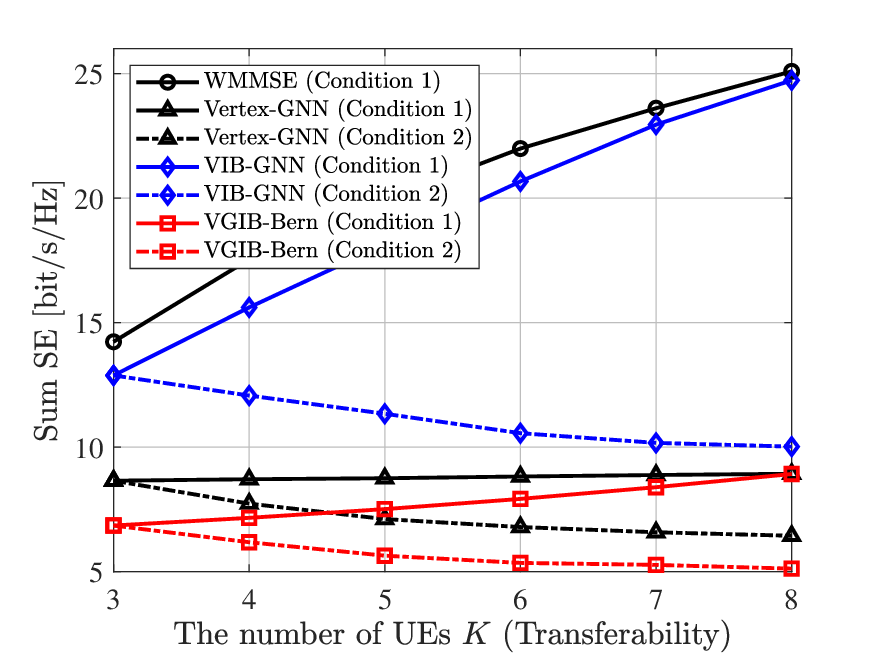}
    \caption{Sum SE against the number of UEs $K$ over different frameworks with $M=10$, $N=4$, and $\sigma_\mathrm{i}^2=0.1$ (Vertex-Transferability).}
\end{figure}
\begin{figure}[t]
\centering
    \includegraphics[scale=0.45]{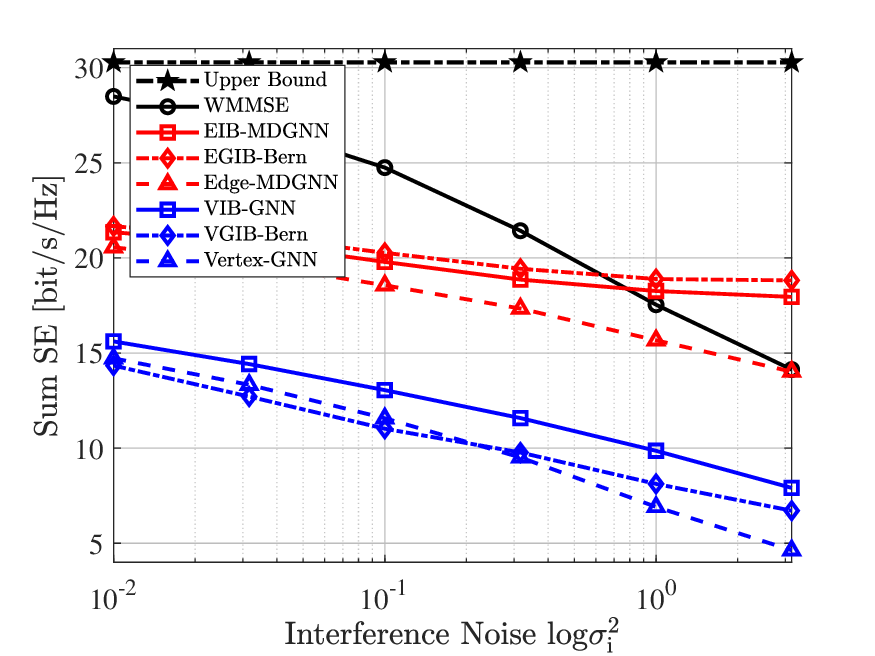}
    \caption{Sum SE against interference noise $\sigma_\mathrm{i}^2$ under ZF precoding over different frameworks with $M=10$, $K=4$, and $N=4$ (Case 2).}
\end{figure}
\begin{figure}[t]
\centering
    \includegraphics[scale=0.45]{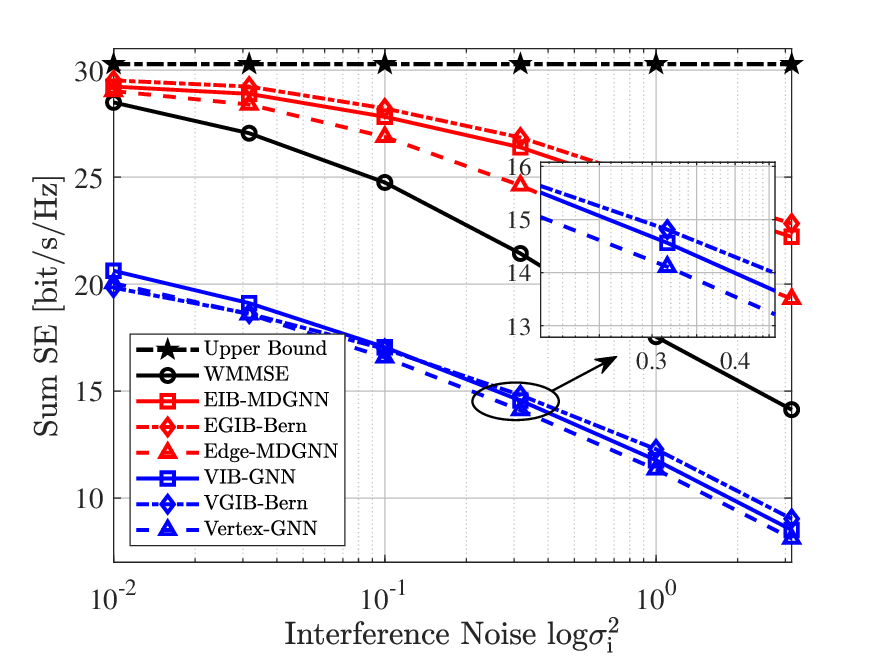}
    \caption{Sum SE against interference noise $\sigma_\mathrm{i}^2$ under L-MMSE precoding over different frameworks with $M=10$, $K=4$, and $N=4$ (Case 2).}
\end{figure}
\emph{Experimental Setup:} To further evaluate the performance of the proposed robust optimization framework on power control problems, we still conduct one set of experiments in the cell-free mMIMO system. Note that the optimization object $p_{mk}$ under case 2 belongs to a 2D architecture, which is different from 3D optimization object $\mathbf{w}_{mk}$ and is independent of antenna variables.
Then, the complete power control problem with the maximum transmission power $p_m$ can be written as
\begin{subequations}
\begin{align}
\!\max_{p_{mk}} \sum_{k=1}^{K}{\text{\rm log}}_{2}&\bigg(1+\frac{\sum_{m=1}^Mp_{mk}\mathbf{E}_{m,kk}}{\sum_{i=1,i\neq k}^K\sum_{m=1}^Mp_{mi}\mathbf{E}_{m,ki} + \sigma^2}\bigg)\\
\mathrm{s.t.}  \qquad\quad\thinspace\thinspace\thinspace &\qquad\quad\thinspace\thinspace\sum_{i=1}^Kp_{mi} \leqslant p_m, \forall m, i,
\end{align}
\end{subequations}
where the simulation parameter settings in case 2 are the same as those in case 1.

\emph{Permutation Properties:} Regarding the power control optimization problem $\mathbf{p}_2=\mathcal{G}_2(\mathbf{h}_2)$ with vectorized channel information $\mathbf{h}_2 \in \mathbb{C}^{MK\times N}$ and optimization variable $\mathbf{p}_2 \in \mathbb{C}^{MK}$, which belongs to a permutation problem of AP-set $\mathcal{M}$ and UE-set $\mathcal{K}$ in non-nested sets, with antenna-set being non-permutation, satisfying 2D-PE property and 1D-PE property with fewer trainable parameters as
\begin{subequations}
\begin{align}
\mathbf{\Pi}_1^T\mathbf{p}_1&=\mathcal{G}_1(\mathbf{\Pi}_1^T\mathbf{h}_1),\\
(\mathbf{\Pi}_1^T \otimes \mathbf{\Pi}_2^T)\mathbf{p}_2&=\mathcal{G}_2((\mathbf{\Pi}_1^T \otimes \mathbf{\Pi}_2^T)\mathbf{h}_2),
\end{align}
\end{subequations}
where 1D-non-nested permutation $\mathbf{\Omega}_1^{\text{\rm non}} \triangleq \mathbf{\Pi}_1^T$ for any single set with $\mathbf{p}_1\in \mathbb{C}^{d_1 \times d_2}$, $\mathbf{h}_1 \in \mathbb{C}^{d_1 \times d_2 \times N}$, and 2D-non-nested permutation $\mathbf{\Omega}_2^{\text{\rm non}} \triangleq \mathbf{\Pi}_1^T \otimes \mathbf{\Pi}_2^T$ for the joint set. Similarly, the elements in each set can be permuted arbitrarily, and we also can transform the power control problem (27) under GIB into
\begin{subequations}
\begin{align}
\min_{\mathbb{P}(\mathbf{z}_\text{\rm v}^l|\mathcal{D}_{\text{\rm v}})\in \Omega_{\text{\rm v}}}
\!\!\!\!\!\text{VGIB}(\mathcal{D}_{\text{\rm v}},\mathbf{p};\mathbf{z}_\text{\rm v}^l) \!&=\! -I(\mathbf{p};\mathbf{z}_{\text{\rm v}}^L) + \beta I(\mathcal{D}_{\text{\rm v}};\mathbf{z}_{\text{\rm v}}^L),\\
\min_{\mathbb{P}(\mathbf{z}_\text{\rm e}^l|\mathcal{D}_{\text{\rm e}})\in \Omega_{\text{\rm e}}}
\!\!\!\!\!\text{EGIB}(\mathcal{D}_{\text{\rm e}},\mathbf{p};\mathbf{z}_\text{\rm e}^l) \!&=\! -I(\mathbf{p};\mathbf{z}_{\text{\rm e}}^L) + \beta I(\mathcal{D}_{\text{\rm e}};\mathbf{z}_{\text{\rm e}}^L).
\end{align}
\end{subequations}

\emph{Results and Analysis:} We further validate the effectiveness and robustness of the proposed optimization framework in another power control task. Fig. 12 and Fig. 13 show the sum SE performance against the interference noise $\sigma_{\text{\rm i}}^2$ based on two different precoding schemes with $M = 10$, $K = 4$, and $N = 4$, including zero-forcing (ZF) precoding and local minimum mean-square error (L-MMSE) precoding. For ZF precoding shown in Fig. 10, we observe that the proposed EIB-MDGNN and EGIB-Bern undoubtedly outperform traditional GNN frameworks. This is because they utilize hyper-edges to update hidden representations and suppress the aggregation of irrelevant information from neighbors, ensuring the effective suppression of interference using the minimal sufficient information. However, due to the lack of consideration for potential channel noise in ZF precoding, there is a certain sum SE performance loss compared to WMMSE. As for L-MMSE precoding, compared with Fig. 12, the proposed EIB-MDGNN and EGIB-Bern can significantly outperform WMMSE in suppressing interference by thoroughly combining channel noise, and the performance gap between them gradually widens with the increase of the interference noise, e.g., from 2.59\% ($\sigma_{\text{\rm i}}^2=10^{-2}$) to 57.18\% ($\sigma_{\text{\rm i}}^2=10^{0.5}$). This reveals that our proposed frameworks have superior robustness in environments prone to interference. Moreover, comparing the two cases of precoding and power control, as shown in Fig. 6 and Fig. 12, we can clearly observe that under low interference noise, the performance gap between ``regular" GNNs and our proposed frameworks is reduced to within 40\%, which is much smaller than when learning precoding policies. This further indicates that our proposed frameworks have a more significant advantage in learning wireless policies related to channel matrices compared to ``regular" GNNs.
\section{Conclusion}
In this paper, we first introduced an efficient MDGNN framework that effectively utilizes hyper-edges to update hidden representations instead of traditional vertices, significantly reducing information loss during message passing. Then, we proposed a robust optimization framework of MDGNNs with EGIB for signal processing in wireless communications. This demonstrates an innovative optimal hidden representation that provides maximum information about outputs and prevents irrelevant information from being obtained from inputs.
Moreover, we applied the proposed optimization framework to two typical signal processing tasks to verify robustness and effectiveness, including joint precoding and power control. The simulation results showed that the proposed EIB-MDGNN and EGIB-Bern have superior robustness in environments prone to interference. In future work, we aim to extend the consideration of centralized graphs to distributed subgraphs to further meet the requirements of practical deployments.

\begin{appendices}
\section{Proof of Theorem 1}
We apply variational bounds of mutual information $I_\text{\rm{NWJ}}$ proposed by \cite{[31]}, which can be expressed as
\begin{lemm}
For any two variables $X$ and $Y$, and any permutation invariant function $\mathcal{F}(\mathbf{x},\mathbf{y}) \in \mathbb{R}$, we have
\begin{equation}
\setcounter{equation}{30}
\begin{aligned}
I(\mathbf{x};\mathbf{y}) \geqslant \mathbb{E}_{\mathbb{P}(\mathbf{x},\mathbf{y})}[\mathcal{F}(\mathbf{x},\mathbf{y})]- e^{-1}\mathbb{E}_{\mathbb{P}(\mathbf{x})\mathbb{P}(\mathbf{y})}\Big[e^{\mathcal{F}(\mathbf{x},\mathbf{y})}\Big].\!\!
\end{aligned}
\end{equation}
\end{lemm}
We use the above \textbf{Lemma 1} to term $I(\mathbf{y};\mathbf{z}_{\text{\rm v}}^L)$ and plug in
\begin{equation}
\setcounter{equation}{31}
\begin{aligned}
\mathcal{F}(\mathbf{y};\mathbf{z}_{\text{\rm v}}^L) = 1 + \text{\rm{log}}\frac{\prod_{v\in \mathcal{V}}\mathbb{Q}_1(\mathbf{y}_v|\mathbf{z}_{\text{\rm v},v}^L)}{\mathbb{Q}_2(\mathbf{y})}.
\end{aligned}
\end{equation}

Then, term $I(\mathbf{y};\mathbf{z}_{\text{\rm v}}^L)$ can be derived as
\begin{equation}
\setcounter{equation}{32}
\begin{aligned}
I(\mathbf{y};\mathbf{z}_{\text{\rm v}}^L) &\geqslant \mathbb{E}_{\mathbb{P}(\mathbf{y};\mathbf{z}_{\text{\rm v}}^L)}\left[1 + \text{\rm{log}}\frac{\prod_{v\in \mathcal{V}}\mathbb{Q}_1(\mathbf{y}_v|\mathbf{z}_{\text{\rm v},v}^L)}{\mathbb{Q}_2(\mathbf{y})}\right]\\
&\quad- e^{-1}\mathbb{E}_{\mathbb{P}(\mathbf{y})\mathbb{P}(\mathbf{z}_{\text{\rm v}}^L)}\left[e\, \frac{\prod_{v\in \mathcal{V}}\mathbb{Q}_1(\mathbf{y}_v|\mathbf{z}_{\text{\rm v},v}^L)}{\mathbb{Q}_2(\mathbf{y})}\right]\\
&\geqslant 1 + \mathbb{E}_{\mathbb{P}(\mathbf{y};\mathbf{z}_{\text{\rm v}}^L)}\left[\text{\rm{log}}\frac{\prod_{v\in \mathcal{V}}\mathbb{Q}_1(\mathbf{y}_vv|\mathbf{z}_{\text{\rm v},v}^L)}{\mathbb{Q}_2(\mathbf{y})}\right]\\
&\quad- \mathbb{E}_{\mathbb{P}(\mathbf{y})\mathbb{P}(\mathbf{z}_{\text{\rm v}}^L)}\left[\frac{\prod_{v\in \mathcal{V}}\mathbb{Q}_1(\mathbf{y}_v|\mathbf{z}_{\text{\rm v},v}^L)}{\mathbb{Q}_2(\mathbf{y})}\right].
\end{aligned}
\end{equation}
\section{Proof of Theorem 2}
Firstly, we apply the data processing inequality and the Markovian dependency to prove the first inequality.
\begin{lemm}
For any three variables $\mathbf{x}$, $\mathbf{y}$ and $\mathbf{z}$, which follow the Markov Chain $<\mathbf{x}\rightarrow \mathbf{y} \rightarrow \mathbf{z}>$, we have
\begin{equation}
\setcounter{equation}{33}
\begin{aligned}
I(\mathbf{x};\mathbf{y}) \geqslant I(\mathbf{x};\mathbf{z}).
\end{aligned}
\end{equation}
\end{lemm}
We directly apply \textbf{Lemma 2} to the Markov Chain in VGIB, i.e., $<\mathcal{D}_{\text{\rm v}}\rightarrow \{\{\mathbf{z}_{A}^l\}_{l \in \mathcal{S}_A} \cup \{\mathbf{z}_{\text{\rm v}}^l\}_{l \in \mathcal{S}_{\text{\rm v}}}\} \rightarrow \mathbf{z}_{\text{\rm v}}^L>$, which satisfies the Markov property $\mathcal{D}_{\text{\rm v}} \bot \mathbf{z}_{\text{\rm v}}^L|\{\{\mathbf{z}_{A}^l\}_{l \in \mathcal{S}_A} \cup \{\mathbf{z}_{\text{\rm v}}^l\}_{l \in \mathcal{S}_{\text{\rm v}}}\}$ as
\begin{equation}
\setcounter{equation}{34}
\begin{aligned}
I(\mathcal{D}_{\text{\rm v}};\mathbf{z}_{\text{\rm v}}^L) \leqslant I(\mathcal{D}_{\text{\rm v}};\{\mathbf{z}_{A}^l\}_{l \in \mathcal{S}_A} \cup \{\mathbf{z}_{\text{\rm v}}^l\}_{l \in \mathcal{S}_{\text{\rm v}}}).
\end{aligned}
\end{equation}

Then, to prove the second inequality, we define an order ``$\prec$" of random variables in $\{\mathbf{z}_{A}^l\}_{l \in \mathcal{S}_A} \cup \{\mathbf{z}_{\text{\rm v}}^l\}_{l \in \mathcal{S}_{\text{\rm v}}}$. Based on the order $\mathbf{z}_{A}^l \prec \mathbf{z}_{\text{\rm v}}^l$, we define a sequence of sets $H_{A}^l$ and $H_{\text{\rm v}}^l$ as follows:
\begin{equation}
\setcounter{equation}{35}
H_{A}^l= \left\{\mathbf{z}_{\text{\rm v}}^{l_1},\mathbf{z}_{A}^{l_2}|l_1<l,l_2<l,l_1\in \mathcal{S}_{\text{\rm v}},l_2\in \mathcal{S}_A\right\},
\end{equation}
\begin{equation}
\setcounter{equation}{36}
H_{\text{\rm v}}^l= \left\{\mathbf{z}_{\text{\rm v}}^{l_1},\mathbf{z}_{A}^{l_2}|l_1<l,l_2\leqslant l,l_1\in \mathcal{S}_{\text{\rm v}},l_2\in \mathcal{S}_A\right\}.
\end{equation}

Then, we can decompose $I(\mathcal{D}_{\text{\rm v}};\{\mathbf{z}_{A}^l\}_{l \in \mathcal{S}_A} \cup \{\mathbf{z}_{\text{\rm v}}^l\}_{l \in \mathcal{S}_{\text{\rm v}}})$ with respect to this order
\begin{equation}
\setcounter{equation}{37}
\begin{aligned}
&I(\mathcal{D}_{\text{\rm v}};\{\mathbf{z}_{A}^l\}_{l \in \mathcal{S}_A} \cup \{\mathbf{z}_{\text{\rm v}}^l\}_{l \in \mathcal{S}_{\text{\rm v}}}) \\
&=\sum_{l\in \mathcal{S}_A} I(\mathcal{D}_{\text{\rm v}};\mathbf{z}_{A}^l|H_{A}^l)+ \sum_{l\in \mathcal{S}_{\text{\rm v}}} I(\mathcal{D}_{\text{\rm v}};\mathbf{z}_{\text{\rm v}}^l|H_{\text{\rm v}}^l).
\end{aligned}
\end{equation}

\begin{lemm}
For any two variables $\mathbf{x}$ and $\mathbf{y}$, we have the variational upper bound of term $I(\mathbf{x};\mathbf{y})$ with non-negativity of KL divergence as
\begin{equation}
\setcounter{equation}{38}
\begin{aligned}
I(\mathbf{x};\mathbf{y}) &= \mathbb{E}_{\mathbb{P}(\mathbf{x};\mathbf{y})}\left[\text{\rm log}\frac{\mathbb{P}(\mathbf{y}|\mathbf{x})}{\mathbb{P}(\mathbf{y})}\right]\\
&=\mathbb{E}_{\mathbb{P}(\mathbf{x},\mathbf{y})}\left[\text{\rm log}\frac{\mathbb{P}(\mathbf{y}|\mathbf{x})\mathbb{Q}(\mathbf{y})}{\mathbb{P}(\mathbf{y})\mathbb{Q}(\mathbf{y})}\right]\\
&=\mathbb{E}_{\mathbb{P}(\mathbf{x},\mathbf{y})}\left[\text{\rm log}\frac{\mathbb{P}(\mathbf{y}|\mathbf{x})}{\mathbb{Q}(\mathbf{y})}\right]-\mathbb{E}_{\mathbb{P}(\mathbf{y})}\left[\text{\rm log}\frac{\mathbb{P}(\mathbf{y})}{\mathbb{Q}(\mathbf{y})}\right]\\
&=\mathbb{E}_{\mathbb{P}(\mathbf{x},\mathbf{y})}\left[\text{\rm log}\frac{\mathbb{P}(\mathbf{y}|\mathbf{x})}{\mathbb{Q}(\mathbf{y})}\right]-\text{\rm KL}(\mathbb{P}(\mathbf{y})\parallel\mathbb{Q}(\mathbf{y}))\\
&\leqslant\mathbb{E}_{\mathbb{P}(\mathbf{x},\mathbf{y})}\left[\text{\rm log}\frac{\mathbb{P}(\mathbf{y}|\mathbf{x})}{\mathbb{Q}(\mathbf{y})}\right].
\end{aligned}
\end{equation}
\end{lemm}

Correspondingly, we can bound first term $I(\mathcal{D}_{\text{\rm v}};\mathbf{z}_{A}^l|H_{A}^l)$ in equation (37) with the above \textbf{Lemma 3} as
\begin{equation}
\setcounter{equation}{39}
\begin{aligned}
I(\mathcal{D}_{\text{\rm v}};\mathbf{z}_{A}^l|H_{A}^l) &\overset{1)}{\leqslant} I(\mathcal{D}_{\text{\rm v}},\mathbf{z}_{\text{\rm v}}^{l-1};\mathbf{z}_{A}^l|H_{A}^l)\\
&\overset{2)}{=} I(\mathbf{A}_{\text{\rm v}},\mathbf{z}_{\text{\rm v}}^{l-1};\mathbf{z}_{A}^l|H_{A}^l)\\
&\quad+ I(\mathbf{x}_{\text{\rm v}};\mathbf{z}_{A}^l|H_{A}^l,\mathbf{A}_{\text{\rm v}},\mathbf{z}_{\text{\rm v}}^{l-1})\\
&\overset{3)}{=} I(\mathbf{A}_{\text{\rm v}},\mathbf{z}_{\text{\rm v}}^{l-1};\mathbf{z}_{A}^l|H_{A}^l)\\
&\overset{4)}{\leqslant} I(\mathbf{A}_{\text{\rm v}},\mathbf{z}_{\text{\rm v}}^{l-1};\mathbf{z}_{A}^l)\overset{5)}{\leqslant}\text{\rm A-VGIB}^l,
\end{aligned}
\end{equation}
where
\begin{equation}
\setcounter{equation}{40}
\text{\rm{A-VGIB}}^l=\mathbb{E}_{\mathbb{P}(\mathbf{z}_{A}^l,\mathbf{A}_{\text{\rm v}},\mathbf{z}_{\text{\rm v}}^{l-1})}\left[\text{\rm{log}}\frac{\mathbb{P}(\mathbf{z}_{A}^l|\mathbf{A}_{\text{\rm v}},\mathbf{z}_{\text{\rm v}}^{l-1})}{\mathbb{Q}(\mathbf{z}_{A}^l)}\right],
\end{equation}
and 1), 2) use the basic properties of mutual information; 3) uses $\mathbf{x}_{\text{\rm v}} \bot \mathbf{z}_{A}^l |\{\mathbf{A}_{\text{\rm v}},\mathbf{z}_{\text{\rm v}}^{l-1}\}$ with $I(\mathbf{x}_{\text{\rm v}};\mathbf{z}_{A}^l|H_{A}^l,\mathbf{A}_{\text{\rm v}},\mathbf{z}_{\text{\rm v}}^{l-1})=0$; 4) uses $H_A^l \bot \mathbf{z}_{A}^l | \{\mathbf{A}_{\text{\rm v}},\mathbf{z}_{\text{\rm v}}^{l-1}\}$; and 5) uses $\textbf{Lemma 3}$.

Similarly, we have second term $I(\mathcal{D}_{\text{\rm v}};\mathbf{z}_{\text{\rm v}}^{l}|H_{\text{\rm v}}^l)$ as
\begin{equation}
\setcounter{equation}{41}
\begin{aligned}
I(\mathcal{D}_{\text{\rm v}};\mathbf{z}_{\text{\rm v}}^{l}|H_{\text{\rm v}}^l) &\overset{1)}{\leqslant} I(\mathcal{D}_{\text{\rm v}},\mathbf{z}_{\text{\rm v}}^{l-1},\mathbf{z}_{A}^l;\mathbf{z}_{\text{\rm v}}^{l}|H_{\text{\rm v}}^l)\\
&\overset{2)}{=} I(\mathbf{z}_{\text{\rm v}}^{l-1},\mathbf{z}_{A}^l;\mathbf{z}_{\text{\rm v}}^{l}|H_{\text{\rm v}}^l)\\
&\quad+I(\mathcal{D}_{\text{\rm v}};\mathbf{z}_{\text{\rm v}}^{l}|H_{\text{\rm v}}^l,\mathbf{z}_{\text{\rm v}}^{l-1},\mathbf{z}_{A}^l)\\
&\overset{3)}{=} I(\mathbf{z}_{\text{\rm v}}^{l-1},\mathbf{z}_{A}^l;\mathbf{z}_{\text{\rm v}}^{l}|H_{\text{\rm v}}^l)\\
&\overset{4)}{\leqslant} I(\mathbf{z}_{\text{\rm v}}^{l-1},\mathbf{z}_{A}^l;\mathbf{z}_{\text{\rm v}}^{l})\overset{5)}{\leqslant}\text{\rm V-VGIB}^l,
\end{aligned}
\end{equation}
where
\begin{equation}
\setcounter{equation}{42}
\text{\rm{V-VGIB}}^l=\mathbb{E}_{\mathbb{P}(\mathbf{z}_{\text{\rm v}}^{l},\mathbf{z}_{\text{\rm v}}^{l-1},\mathbf{z}_{A}^l)}\left[\text{\rm{log}}\frac{\mathbb{P}(\mathbf{z}_{\text{\rm v}}^{l}|\mathbf{z}_{\text{\rm v}}^{l-1},\mathbf{z}_{A}^l)}{\mathbb{Q}(\mathbf{z}_{\text{\rm v}}^{l})}\right],
\end{equation}
and 1), 2) use the basic properties of mutual information; 3) uses $\mathcal{D}_{\text{\rm v}} \bot \mathbf{z}_{\text{\rm v}}^{l} |\{\mathbf{z}_{\text{\rm v}}^{l-1},\mathbf{z}_{A}^l\}$ with $I(\mathcal{D}_{\text{\rm v}};\mathbf{z}_{\text{\rm v}}^{l}|H_{\text{\rm v}}^l,\mathbf{z}_{\text{\rm v}}^{l-1},\mathbf{z}_{A}^l)=0$; 4) uses $H_{\text{\rm v}}^l \bot \mathbf{z}_{\text{\rm v}}^{l} | \{\mathbf{z}_{\text{\rm v}}^{l-1},\mathbf{z}_{A}^l\}$; and 5) uses $\textbf{Lemma 3}$. and 5) uses $\textbf{Lemma 3}$. and 5) uses $\textbf{Lemma 3}$. and 5) uses $\textbf{Lemma 3}$. and 5) uses $\textbf{Lemma 3}$. and 5) uses $\textbf{Lemma 3}$.

\section{Proof of Corollary 1}
Similarly, we apply the variational bounds of mutual information $I_\text{\rm{NWJ}}$ proposed by \cite{[31]}, and we adopt the above \textbf{Lemma 1} to term $I(\mathbf{y};\mathbf{z}_{\text{\rm e}}^L)$ and plug in
\begin{equation}
\setcounter{equation}{43}
\begin{aligned}
\mathcal{F}(\mathbf{y};\mathbf{z}_{\text{\rm e}}^L) = 1 + \text{\rm{log}}\frac{\prod_{i\in \mathcal{N}_{\text{\rm e}}}\prod_{e_i\in \mathcal{E}_i}\mathbb{Q}_1(\mathbf{y}_{e_i}|\mathbf{z}_{{\text{\rm e}},e_i}^L)}{\mathbb{Q}_2(\mathbf{y})}.
\end{aligned}
\end{equation}

Then, term $I(\mathbf{y};\mathbf{z}_{\text{\rm e}}^L)$ can be derived as
\begin{equation}
\setcounter{equation}{44}
\begin{aligned}
I(\mathbf{y};\mathbf{z}_{\text{\rm e}}^L)
&\geqslant 1 + \mathbb{E}_{\mathbb{P}(\mathbf{y};\mathbf{z}_{\text{\rm e}}^L)}\left[\text{\rm{log}}\frac{\prod_{i\in \mathcal{N}_{\text{\rm e}}}\prod_{{e_i}\in \mathcal{E}_i}\mathbb{Q}_1(\mathbf{y}_{e_i}|\mathbf{z}_{{\text{\rm e}},e_i}^L)}{\mathbb{Q}_2(\mathbf{y})}\right]\\
&\quad- \mathbb{E}_{\mathbb{P}(\mathbf{y})\mathbb{P}(\mathbf{z}_{\text{\rm e}}^L)}\left[\frac{\prod_{i\in \mathcal{N}_{\text{\rm e}}}\prod_{{e_i}\in \mathcal{E}_i}\mathbb{Q}_1(\mathbf{y}_{e_i}|\mathbf{z}_{{\text{\rm e}},e_i}^L)}{\mathbb{Q}_2(\mathbf{y})}\right].
\end{aligned}
\end{equation}
\vspace{-0.7cm}
\section{Proof of Corollary 2}
Similarly, based on \textbf{Lemma 2}, we can apply it to the Markov Chain in EGIB, i.e., $<\mathcal{D}_{\text{\rm e}}\rightarrow \{\{\cup_{i \in \mathcal{N}_{\text{\rm e}}}\{\mathbf{z}_{A_i}^{l}\}_{l \in \mathcal{S}_{A_i}}\} \cup \{\mathbf{z}_{\text{\rm e}}^l\}_{l \in \mathcal{S}_{\text{\rm e}}}\} \rightarrow \mathbf{z}_{\text{\rm e}}^L>$, which satisfies the Markov property $\mathcal{D}_{\text{\rm e}} \bot \mathbf{z}_{\text{\rm e}}^L|\{\{\cup_{i \in \mathcal{N}_{\text{\rm e}}}\{\mathbf{z}_{A_i}^{l}\}_{l \in \mathcal{S}_{A_i}}\} \cup \{\mathbf{z}_{\text{\rm e}}^l\}_{l \in \mathcal{S}_{\text{\rm e}}}\}$. Then, we can prove the first inequality in (16) as
\begin{equation}
\setcounter{equation}{45}
\begin{aligned}
I(\mathcal{D}_{\text{\rm e}};Z_{\text{\rm e}}^L) \leqslant  I(\mathcal{D}_{\text{\rm e}};\{\cup_{i \in \mathcal{N}_{\text{\rm e}}}\{\mathbf{z}_{A_i}^{l}\}_{l \in \mathcal{S}_{A_i}}\} \cup \{\mathbf{z}_{\text{\rm e}}^l\}_{l \in \mathcal{S}_{\text{\rm e}}}).
\end{aligned}
\end{equation}

Moreover, to prove the second inequality, we define an order ``$\prec$" of random variables in $\{\cup_{i \in \mathcal{N}_{\text{\rm e}}}\{\mathbf{z}_{A_i}^{l}\}_{l \in \mathcal{S}_{A_i}}\} \cup \{\mathbf{z}_{\text{\rm e}}^l\}_{l \in \mathcal{S}_{\text{\rm e}}}$. Based on the order, we define a sequence of sets as follows:
\begin{subequations}
\begin{align}
H_{A_i}^{l}&= \left\{\mathbf{z}_{\text{\rm e}}^{l_2},\mathbf{z}_{A_1}^{l_3},\ldots,\mathbf{z}_{A_{\mathcal{N}_{\text{\rm e}}}}^{l_{{\mathcal{N}_{\text{\rm e}}}+2}}|l_2,l_{j_1}<l,l_{j_2}\leqslant l\right\},\\
H_{\text{\rm e}}^{l}&= \left\{\mathbf{z}_{\text{\rm e}}^{l_2},\mathbf{z}_{A_1}^{l_3},\ldots,\mathbf{z}_{A_{\mathcal{N}_{\text{\rm e}}}}^{l_{{\mathcal{N}_{\text{\rm e}}}+2}}|l_2<l,l_j \leqslant l.\right\},
\end{align}
\end{subequations}
where $l_2\in \mathcal{S}_{\text{\rm e}},l_{j_1}\in \mathcal{S}_{A_{j_1-2}},l_{j_2}\in \mathcal{S}_{A_{j_2-2}},l_{j}\in \mathcal{S}_{A_{j-2}},j_1\in \{i+2,\ldots,{\mathcal{N}_{\text{\rm e}}}+2\},j_2\in \{3,\ldots,i+1\},j\in \{3,\ldots,{\mathcal{N}_{\text{\rm e}}}+2\}.$
Then, we also can decompose $I(\mathcal{D}_{\text{\rm e}};\{\cup_{i \in \mathcal{N}_{\text{\rm e}}}\{\mathbf{z}_{A_i}^{l_i}\}_{l_i \in \mathcal{S}_{A_i}}\} \cup \{\mathbf{z}_{\text{\rm e}}^l\}_{l \in \mathcal{S}_{\text{\rm e}}})$ with respect to this order
\begin{equation}
\setcounter{equation}{47}
\begin{aligned}
&I(\mathcal{D}_{\text{\rm e}};\{\cup_{i \in \mathcal{N}_{\text{\rm e}}}\{\mathbf{z}_{A_i}^{l}\}_{l \in \mathcal{S}_{A_i}}\} \cup \{\mathbf{z}_{\text{\rm e}}^l\}_{l \in \mathcal{S}_{\text{\rm e}}}) \\
&=\sum_{i\in \mathcal{N}_{\text{\rm e}}}\sum_{l\in \mathcal{S}_{A_i}} I(\mathcal{D}_{\text{\rm e}};\mathbf{z}_{A_i}^{l}|H_{A_i}^{l}) + \sum_{l\in \mathcal{S}_{\text{\rm e}}} I(\mathcal{D}^{\mathrm{e}};\mathbf{z}_{\text{\rm e}}^l|H_{\text{\rm e}}^l).
\end{aligned}
\end{equation}

Correspondingly, we can bound first term $I(\mathcal{D}_{\text{\rm e}};\mathbf{z}_{A_1}^{l}|H_{A_1}^{l})$ in equation (47) with respect to $i=1$ and $i \geqslant 2$ as
\begin{equation}
\setcounter{equation}{48}
\begin{aligned}
I(\mathcal{D}_{\text{\rm e}};\mathbf{z}_{A_1}^{l}|H_{A_1}^{l}) &\overset{1)}{\leqslant} I(\mathcal{D}_{\text{\rm e}},\mathbf{z}_{\text{\rm e}}^{l-1};\mathbf{z}_{\mathbf{A}_{\text{\rm e},1}}^{l}|H_{A_1}^{l})\\
&\overset{2)}{=} I(\mathbf{A}_{\text{\rm e},1},\mathbf{z}_{\text{\rm e}}^{l-1};\mathbf{z}_{A_1}^{l}|H_{A_1}^{l})\\
+\, &I(\mathcal{A}_e\backslash \mathbf{A}_{\text{\rm e},1},\mathbf{x}_{\text{\rm e}};\mathbf{z}_{A_1}^{l}|H_{A_1}^{l},\mathbf{A}_{\text{\rm e},1},\mathbf{z}_{\text{\rm e}}^{l-1})\\
&\overset{3)}{=} I(\mathbf{A}_{\text{\rm e},1},\mathbf{z}_{\text{\rm e}}^{l-1};\mathbf{z}_{A_1}^{l}|H_{A_1}^{l})\\
&\overset{4)}{\leqslant} I(\mathbf{A}_{\text{\rm e},1},\mathbf{z}_{\text{\rm e}}^{l-1};Z_{A_1}^{l})\overset{5)}{\leqslant}\text{\rm{A-EGIB}}_1^{l},
\end{aligned}
\end{equation}
\begin{equation}
\setcounter{equation}{49}
\begin{aligned}
\!I(\mathcal{D}_{\text{\rm e}};\mathbf{z}_{A_i}^{l}|H_{A_i}^{l}) &\overset{1)}{\leqslant} I(\mathcal{D}_{\text{\rm e}},\mathbf{z}_{\text{\rm e}}^{l-1},\mathcal{Z}_{i}^{\text{\rm e}};\mathbf{z}_{A_i}^{l}|H_{A_i}^{l})\\
&\overset{2)}{=} I(\mathbf{A}_{\text{\rm e},i},\mathbf{z}_{\text{\rm e}}^{l-1},\mathcal{Z}_{i}^e;\mathbf{z}_{A_i}^{l}|H_{A_i}^{l})\\
+\, &I(\mathcal{A}_{\text{\rm e}}\backslash \mathbf{A}_{\text{\rm e},i},E;\mathbf{z}_{A_i}^{l}|H_{A_i}^{l},\mathbf{A}_{\text{\rm e},i},\mathbf{z}_{\text{\rm e}}^{l-1},\mathcal{Z}_{i}^{\text{\rm e}})\\
&\overset{3)}{=} I(\mathbf{A}_{\text{\rm e},i},\mathbf{z}_{\text{\rm e}}^{l-1},\mathcal{Z}_{i}^{\text{\rm e}};\mathbf{z}_{A_i}^{l}|H_{A_i}^{l})\\
&\overset{4)}{\leqslant} I(\mathbf{A}_{\text{\rm e},i},\mathbf{z}_{\text{\rm e}}^{l-1},\mathcal{Z}_{i}^e;\mathbf{z}_{A_i}^{l})\overset{5)}{\leqslant}\text{\rm{A-EGIB}}_i^{l},
\end{aligned}
\end{equation}
where $\text{\rm{A-EGIB}}_1^{l}$ and $\text{\rm{A-EGIB}}_i^{l}$ with $\mathcal{Z}_{i}^{\text{\rm e}}=\{\mathbf{z}_{A_1}^{l},\ldots,\mathbf{z}_{A_{i-1}}^{l}\}$, $i \geqslant 2$ satisfy
\begin{equation}
\setcounter{equation}{50}
\begin{aligned}
\!\!\!\!\text{\rm{A-EGIB}}_{1}^{l}= \mathbb{E}_{\mathbb{P}(\mathbf{z}_{A_1}^{l},\mathbf{A}_{\text{\rm e},1},\mathbf{z}_{\text{\rm e}}^{l-1})}\Big[\text{\rm{log}}\frac{\mathbb{P}(\mathbf{z}_{A_1}^{l}|\mathbf{A}_{\text{\rm e},1},\mathbf{z}_{\text{\rm e}}^{l-1})}{\mathbb{Q}(\mathbf{z}_{A_1}^{l})}\Big],
\end{aligned}
\end{equation}
and
\begin{equation}
\setcounter{equation}{51}
\begin{aligned}
\text{\rm{A-EGIB}}_{i}^{l}=\mathbb{E}_{\mathbb{P}(\mathbf{z}_{A_i}^{l},\mathbf{A}_{\text{\rm e},i},\mathbf{z}_{\text{\rm e}}^{l-1},\mathcal{Z}_{i}^{\text{\rm e}})}
\bigg[\text{\rm{log}}\frac{\mathbb{P}(\mathbf{z}_{A_i}^{l}|\mathbf{A}_{\text{\rm e},i},\mathbf{z}_{\text{\rm e}}^{l-1},\mathcal{Z}_{i}^{\text{\rm e}})}{\mathbb{Q}(\mathbf{z}_{A_i}^{l})}\bigg].
\end{aligned}
\end{equation}

Note that 1), 2) use the basic properties of mutual information;
3) use $\{\mathcal{A}_{\text{\rm e}}\backslash \mathbf{A}_{\text{\rm e},1},\mathbf{x}_{\text{\rm e}}\} \bot \mathbf{z}_{A_1}^{l} |\{\mathbf{A}_{\text{\rm e},1},\mathbf{z}_{\text{\rm e}}^{l-1}\}$ with $I(\mathcal{A}_{\text{\rm e}}\backslash \mathbf{A}_{\text{\rm e},1},\mathbf{x}_{\text{\rm e}};\mathbf{z}_{A_1}^{l} |H_{A_1}^{l},\mathbf{A}_{\text{\rm e},1},\mathbf{z}_{\text{\rm e}}^{l-1})=0$ and $\{\mathcal{A}_{\text{\rm e}}\backslash \mathbf{A}_{\text{\rm e},i},\mathbf{x}_{\text{\rm e}}\}$ $ \bot \mathbf{z}_{A_i}^{l}| \{\mathbf{A}_{\text{\rm e},i},\mathbf{z}_{\text{\rm e}}^{l-1},\mathcal{Z}_{i}^{\text{\rm e}}\}$ with $I(\mathcal{A}_{\text{\rm e}}\backslash \mathbf{A}_{\text{\rm e},i},\mathbf{x}_{\text{\rm e}};\mathbf{z}_{A_i}^{l}|$ $H_{A_i}^{l},\mathbf{A}_{\text{\rm e},i},\mathbf{z}_{\text{\rm e}}^{l-1},\mathcal{Z}_{i}^{\text{\rm e}})=0$;
4) uses $H_{A_1}^{l} \bot \mathbf{z}_{A_1}^{l} | \{\mathbf{A}_{\text{\rm e},1},\mathbf{z}_{\text{\rm e}}^{l-1}\}$ and $H_{A_i}^{l} \bot \mathbf{z}_{A_i}^{l} | \{\mathbf{A}_{\text{\rm e},i},\mathbf{z}_{\text{\rm e}}^{l-1},\mathcal{Z}_{i}^{\text{\rm e}}\}$;
and 5) uses $\textbf{Lemma 3}$.

Similarly, we have second term $I(\mathcal{D}_{\text{\rm e}};\mathbf{z}_{\text{\rm e}}^l|H_{\text{\rm e}}^l)$ as
\begin{equation}
\setcounter{equation}{52}
\begin{aligned}
I(\mathcal{D}_{\text{\rm e}};\mathbf{z}_{\text{\rm e}}^l|H_{\text{\rm e}}^l) &\overset{1)}{\leqslant} I(\mathcal{D}_{\text{\rm e}},\mathbf{z}_{\text{\rm e}}^{l-1},\mathcal{Z}_{A}^{\text{\rm e}};\mathbf{z}_{\text{\rm e}}^l|H_{\text{\rm e}}^l)\\
&\overset{2)}{=} I(\mathbf{z}_{\text{\rm e}}^{l-1},\mathcal{Z}_{A}^{\text{\rm e}};\mathbf{z}_{\text{\rm e}}^l|H_{\text{\rm e}}^l)\\
&\quad+ I(\mathcal{D}_{\text{\rm e}};\mathbf{z}_{\text{\rm e}}^l|H_{\text{\rm e}}^l,\mathbf{z}_{\text{\rm e}}^{l-1},\mathcal{Z}_{A}^{\text{\rm e}})\\
&\overset{3)}{=} I(\mathbf{z}_{\text{\rm e}}^{l-1},\mathcal{Z}_{A}^{\text{\rm e}};\mathbf{z}_{\text{\rm e}}^l|H_{\text{\rm e}}^l)\\
&\overset{4)}{\leqslant} I(\mathbf{z}_{\text{\rm e}}^{l-1},\mathcal{Z}_{A}^{\text{\rm e}};\mathbf{z}_{\text{\rm e}}^{l})\overset{5)}{\leqslant}\text{\rm E-EGIB}^l,
\end{aligned}
\end{equation}
where
\begin{equation}
\setcounter{equation}{53}
\begin{aligned}
\text{\rm{E-EGIB}}^l=\mathbb{E}_{\mathbb{P}(\mathbf{z}_{\text{\rm e}}^{l},\mathbf{z}_{\text{\rm e}}^{l-1},\mathcal{Z}_{A}^{\text{\rm e}})}\left[\text{\rm{log}}\frac{\mathbb{P}(\mathbf{z}_{\text{\rm e}}^{l}|\mathbf{z}_{\text{\rm e}}^{l-1},\mathcal{Z}_{A}^{\text{\rm e}})}{\mathbb{Q}(\mathbf{z}_{\text{\rm e}}^{l})}\right]
\end{aligned}
\end{equation}
with $\mathcal{Z}_{A}^{\text{\rm e}}=\{\mathbf{z}_{A_1}^l,\ldots,\mathbf{z}_{A_{{N}_{\text{\rm e}}}}^l\}$, and 1), 2) use the basic properties of mutual information; 3) uses $\mathcal{D}_{\text{\rm e}} \bot \mathbf{z}_{\text{\rm e}}^l |\{\mathbf{z}_{\text{\rm e}}^{l-1},\mathcal{Z}_A^{\text{\rm e}}\}$ with $I(\mathcal{D}_{\text{\rm e}};\mathbf{z}_{\text{\rm e}}^{l}|H_{\text{\rm e}}^l,\mathbf{z}_{\text{\rm e}}^{l-1},\mathcal{Z}_{A}^{\text{\rm e}})=0$; 4) uses $H_{\text{\rm e}}^l \bot \mathbf{z}_{\text{\rm e}}^l | \{\mathbf{z}_{\text{\rm e}}^{l-1},\mathcal{Z}_A^{\text{\rm e}}\}$; and 5) uses $\textbf{Lemma 3}$.
\end{appendices}

\bibliographystyle{IEEEtran}
\bibliography{IEEEabrv,Ref}
\end{document}